\documentclass[aps,showpacs,prb,twocolumn,floatfix,superscriptaddress,floatfix]{revtex4}
\usepackage[english]{babel}
\usepackage{graphicx}
\usepackage{color}
\usepackage[utf8]{inputenc}
\usepackage{pstricks,pst-grad,color}
\usepackage{graphicx,amssymb}
\usepackage{amsmath}
\usepackage{amssymb}
\usepackage{bbold}
\usepackage{placeins}
\usepackage{bm}

\begin{document}

\title{Magnetic states in a three-dimensional topological Kondo insulator}

\begin{abstract}
We theoretically study the magnetic phase diagram of a three-dimensional topological Kondo insulator by means of real-space dynamical mean field theory. We find that ferromagnetically ordered states become stable upon hole doping. Besides a wide ferromagnetic phase, we observe surface magnetism close to half-filling, which corresponds to an A-type antiferromagnetic state.  We further study the impact of the magnetism on the symmetry protected surface states and find that depending on the surface and the magnetization direction, surface states are still protected by reflection symmetry present in our model. The symmetry protected surface states are shifted away by the magnetization from their original high symmetry momenta in the Brillouin zone. Remarkably, due to the magnetization, the surface states are deformed, resulting in the appearance of arcs in the momentum resolved spectrum. 
\end{abstract}

\author{Robert Peters}
\email[]{peters@scphys.kyoto-u.ac.jp}
\affiliation{Department of Physics, Kyoto University, Kyoto 606-8502, Japan}

\author{Tsuneya Yoshida}
\affiliation{Department of Physics, Kyoto University, Kyoto 606-8502, Japan}

\author{Norio Kawakami}
\affiliation{Department of Physics, Kyoto University, Kyoto 606-8502, Japan}

\newcommand*{\tran}{^{\mkern-1.5mu\mathsf{T}}}
\newcommand{\1}{\mbox{1}\hspace{-0.25em}\mbox{l}}
\date{\today}



\maketitle

\section{introduction}
Topology has become a widely used tool in condensed matter physics for predicting and analyzing symmetry protected surface states which include fascinating particles such as Majorana-, Weyl- or Dirac-fermions\cite{RevModPhys.82.3045,RevModPhys.83.1057}. While the influence of topology in noninteracting systems is well understood by now, the interplay between strong correlations and topology is still obscure. Strong correlations are the origin for phenomena which cannot be seen in noninteracting or weakly interacting systems, such as magnetism, unconventional superconductivity or quantum criticality. Naturally, questions arise such as how the symmetry protected surface states change in the presence of strong interactions or under the influence of magnetism, and whether there are new phenomena which can only be observed in strongly interacting topologically nontrivial systems \cite{Pesin2010,PhysRevLett.106.100403,PhysRevLett.107.010401,0953-8984-25-14-143201,PhysRevB.85.125113,PhysRevLett.112.196404,PhysRevB.85.165138}.
 
One remarkable observation in strongly interacting systems is the reduction of the classification of topological phases in the presence of correlations; the classification of topological phases changes due to strong correlations\cite{PhysRevB.81.134509,PhysRevB.83.075102,PhysRevB.95.045127,PhysRevLett.118.147001}. Other interesting examples are so-called topological Kondo insulators  \cite{PhysRevLett.104.106408,PhysRevB.85.045130,annurev-conmatphys-031214-014749,Takimoto2011,PhysRevLett.110.096401,PhysRevLett.115.156405}, which are topologically nontrivial $f$-electron materials including strong correlations in the $f$-orbital. Candidate materials are for example SmB$_6$\cite{Jiang2013,Neupane2013,PhysRevB.88.121102,PhysRevLett.111.216402,PhysRevX.3.041024,Xu2014} or YbB$_{12}$\cite{Hagiwara2016,PhysRevLett.112.016403}. The topologically nontrivial gap is here formed by a hybridization between conduction- ($c$-) electrons and strongly interacting $f$-electrons. Due to the presence of strong interactions in localized orbitals, the Kondo effect and magnetism can often be observed in $f$-electron materials. Thus, these topological Kondo insulators provide an opportunity to study the interplay between topology and phenomena originating in strong correlations. For example, the interplay between Kondo physics and topology results in the  Kondo breakdown, where the behavior of the topological surface states completely changes at finite temperature \cite{PhysRevLett.111.226403,PhysRevB.93.235159,Chang_Po2017}. Furthermore, these materials have created a stir in the condensed matter community because of the observation of quantum oscillations in strong magnetic fields, which contradicts our common knowledge about insulators\cite{Tan287}.

We here analyze another intriguing phenomenon based on the interplay of nontrivial topology and strong correlations, namely magnetism in a three-dimensional (3D) topological Kondo insulator. Recently  a topological phase has been observed in the magnet Co$_2$MnGa\cite{Belopolski2015} which might open a path for generating highly spin-polarized currents. Furthermore, the Kondo insulator SmB$_6$ is known to have a magnetic phase under pressure\cite{PhysRevLett.94.166401,PhysRevB.77.193107,JPSJ.82.123707,PhysRevLett.116.156401,Chang2017}, which might be an A-type antiferromagnetic state. Thus, a study of magnetism in a topological Kondo insulator and its impact on the symmetry protected surface states are highly desired. 

In this paper, we use the real-space dynamical mean field theory, which allows us to analyze the effect of strong correlations in a topologically nontrivial $f$-electron material and study bulk as well as surface properties. Besides a ferromagnetic phase which is stable upon hole-doping, we find an antiferromagnetic surface state close to half-filling. Although the time-reversal symmetry is broken by the magnetic state, surface states are still protected by the reflection symmetry. We demonstrate that the Dirac cones at the surface of the topological Kondo insulator are shifted and deformed by the magnetization. A remarkable effect of the magnetization on the Dirac cones is the emergence of arcs in the spectrum, which appear due to the energetic splitting of different spin directions.
 
 This paper is organized as follows: In the next section, we will introduce the model and shortly explain the method used to analyze magnetic states. This is followed by sections discussing the phase diagram, the bulk properties and the impact of the magnetism on the surface states. A conclusion finishes the paper.

\section{model and method}

For the purpose of describing magnetism in a topological Kondo insulator, we use a Hamiltonian in a three-dimensional (3D) cubic lattice, which includes two  spin-degenerate orbitals. The orbitals correspond to a conduction ($c$) electron band and an $f$-electron band. 
\begin{figure}[t]
\begin{center}
    \includegraphics[width=1\linewidth]{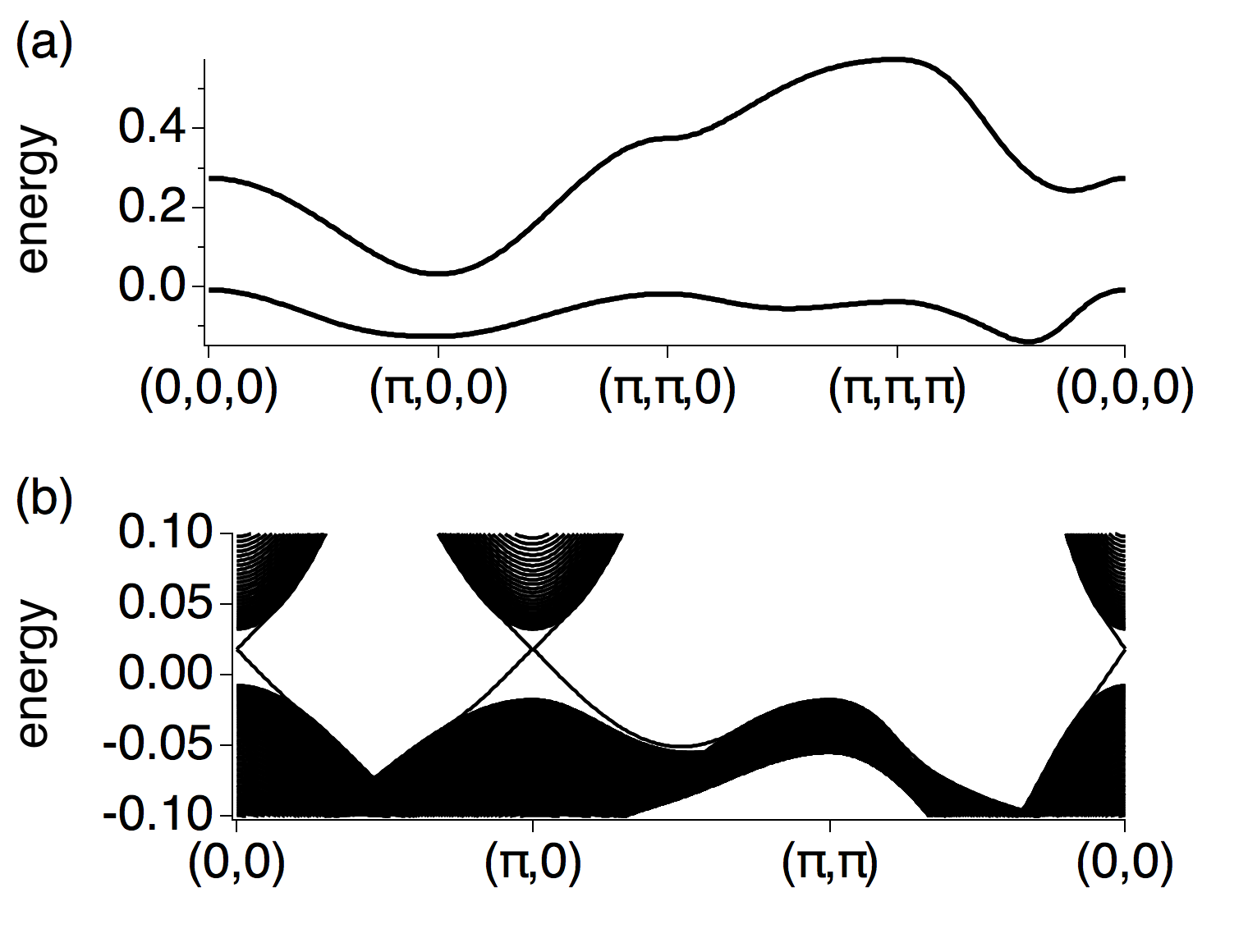}
\end{center}
\caption{(a) Bulk band structure of the noninteracting model for hybridization strength $V=0.06$. (b) Noninteracting band structure for a slab-calculation including $100$ layers with open surfaces. Visible are the Dirac cones at $(k_x,k_y)=(0,0)$ and $(\pi,0)$. Another Dirac cone exists at $(0,\pi)$, which is not shown.
\label{Fig1}}
\end{figure}
The Hamiltonian reads
\begin{eqnarray*}
H&=&H_0+H_{int}\\
H_0&=&\sum_k\sum_{\sigma=\{\uparrow,\downarrow\}}\sum_{o=\{c,f\}}\epsilon^o_kc^\dagger_{k,\sigma,o}c_{k,\sigma,o}\\
&&+V\sum_{k,\tau_1,m\tau_2}c^\dagger_{k,\tau_1,c}c_{k,\tau_2,f}\sin k_x\sigma^x_{\tau_1\tau_2}\\
&&+V\sum_{k,\tau_1,m\tau_2}c^\dagger_{k,\tau_1,c}c_{k,\tau_2,f}\sin k_y\sigma^y_{\tau_1\tau_2}\\
&&+V\sum_{k,\tau_1,m\tau_2}c^\dagger_{k,\tau_1,c}c_{k,\tau_2,f}\sin k_z\sigma^z_{\tau_1\tau_2}\\
&&+0.2\sum_{i,\sigma}n_{i,\sigma,c}\\
\epsilon^c_k&=&-0.1(\cos(k_x)+\cos(k_y)+\cos(k_z))\\
&&+0.075\cos(k_x)\cos(k_y)\\&&+0.075\cos(k_y)\cos(k_z)\\&&+0.075\cos(k_x)\cos(k_z)\\
&&+0.15\cos(k_x)\cos(k_y)\cos(k_z)\\
\epsilon^f_k&=&-0.1\epsilon^c_k\\
H_{int}&=&U\sum_i n_{i,\uparrow,f}n_{i,\downarrow,f}.
\end{eqnarray*}
The operator $c^\dagger_{k,\sigma,o}$ creates an electron with momentum $k$, spin direction $\sigma$ in orbital $o\in\{c,f\}$. $\epsilon^o_k$ describes the energy depending on the momentum for each orbital. The energies have been chosen in a way that there are band inversions between $c$-electrons and $f$-electrons at $(\pi,0,0)$, $(0,\pi,0)$, and $(0,0,\pi)$ in the Brillouin zone, which resembles qualitatively the band structure of SmB$_6$. We include nearest neighbor, next-nearest neighbor and next-next-nearest neighbor hopping on a cubic lattice.
Due to the hybridization, $V$, between the $c$-electron band and the $f$-electron band, a gap opens in the bulk spectrum, see Fig. \ref{Fig1}. We will later use the hybridization strength $V$ as a free parameter in the model. $\sigma^x$, $\sigma^y$, $\sigma^z$ are the Pauli matrices.
The operator  $n_{i,\sigma,c}$ and $n_{i,\sigma,f}$ are local density operators on lattice site $i$ for the $c$-electrons and $f$-electrons, respectively. Finally, $H_{int}$ describes a repulsive local density-density interaction in the $f$-electron band,  necessary to describe the Kondo effect in strongly interacting $f$-electron systems. Throughout this paper we set $U=0.8$.

Because there is an odd number of band inversions between the $c$-electron band and the $f$-electron band, which have different parity, combined with a  gap in the bulk spectrum, this model is a 3D strong topological insulator \cite{PhysRevB.85.045130,PhysRevB.76.045302,PhysRevLett.98.106803,PhysRevLett.104.106408,PhysRevX.2.031008}. The noninteracting band structure with open surfaces, depicted in Fig. \ref{Fig1}(b), shows the surface states at $(k_x,k_y)=(0,0)$ and $(\pi,0)$  on the surface. Another surface state exists at $(k_x,k_y)=(0,\pi)$, which is not shown in Fig. \ref{Fig1}(b).
The inclusion of strong interactions into the $f$-electron band leads to the Kondo effect and a renormalization of the band gap. One remarkable effect of the interaction is the emergence of strongly correlated surface states, which can result in a Kondo breakdown on the surface at finite temperatures \cite{PhysRevLett.111.226403,PhysRevB.93.235159}.

In order to analyze a strongly correlated and topologically nontrivial system with open surfaces, we use the real-space dynamical mean field theory (DMFT). DMFT\cite{Georges1996} maps a lattice model onto a quantum impurity model, which must be solved self-consistently. DMFT thereby includes local fluctuations exactly and is therefore well suited to study the Kondo effect in $f$-electron materials. The real-space DMFT maps each atom of a finite lattice onto a separate quantum impurity model. Thus, the effect of inhomogeneities such as impurities or surfaces can be included into this theory. 

To properly study the above described Hamiltonian, we perform calculations for a homogeneous system, studying bulk properties, and for slabs consisting of $20$ layers. For the homogeneous system we use single-site DMFT focusing on nonmagnetic and ferromagnetic states. The slab calculations are done using the real-space DMFT, where each layer is mapped onto its own quantum impurity model. This provides us the possibility to analyze the impact of the magnetic state on the surface states. Single-site DMFT for bulk ferromagnetism as well as the real-space DMFT calculations are performed self-consistently. 
 For solving the quantum impurity models, we use the numerical renormalization group\cite{Bulla2008}, which is well suited to calculate real-frequency spectral functions and self-energies at low temperatures with high resolution around the Fermi energy\cite{Peters2006,Weichselbaum2007}.

Because we map each layer of our model onto a single quantum impurity model, our ansatz only includes solutions where all atoms in the same layer have the same properties. Thus, we can only describe  in-plane ferromagnetic or paramagnetic (vanishing magnetization) solutions. 
In order to stabilize magnetic states, we dope holes into the $c$-electron band changing the number of $c$-electrons from $n_c=0.9$ to $0.4$.The $f$-electron number is kept fixed at  $n_f=1.1$. Thus, the model is half-filled for $n_c=0.9$. We perform all calculations at $T=0$.

\section{phase diagram}
\begin{figure}[t]
\begin{center}
    \includegraphics[width=0.99\linewidth]{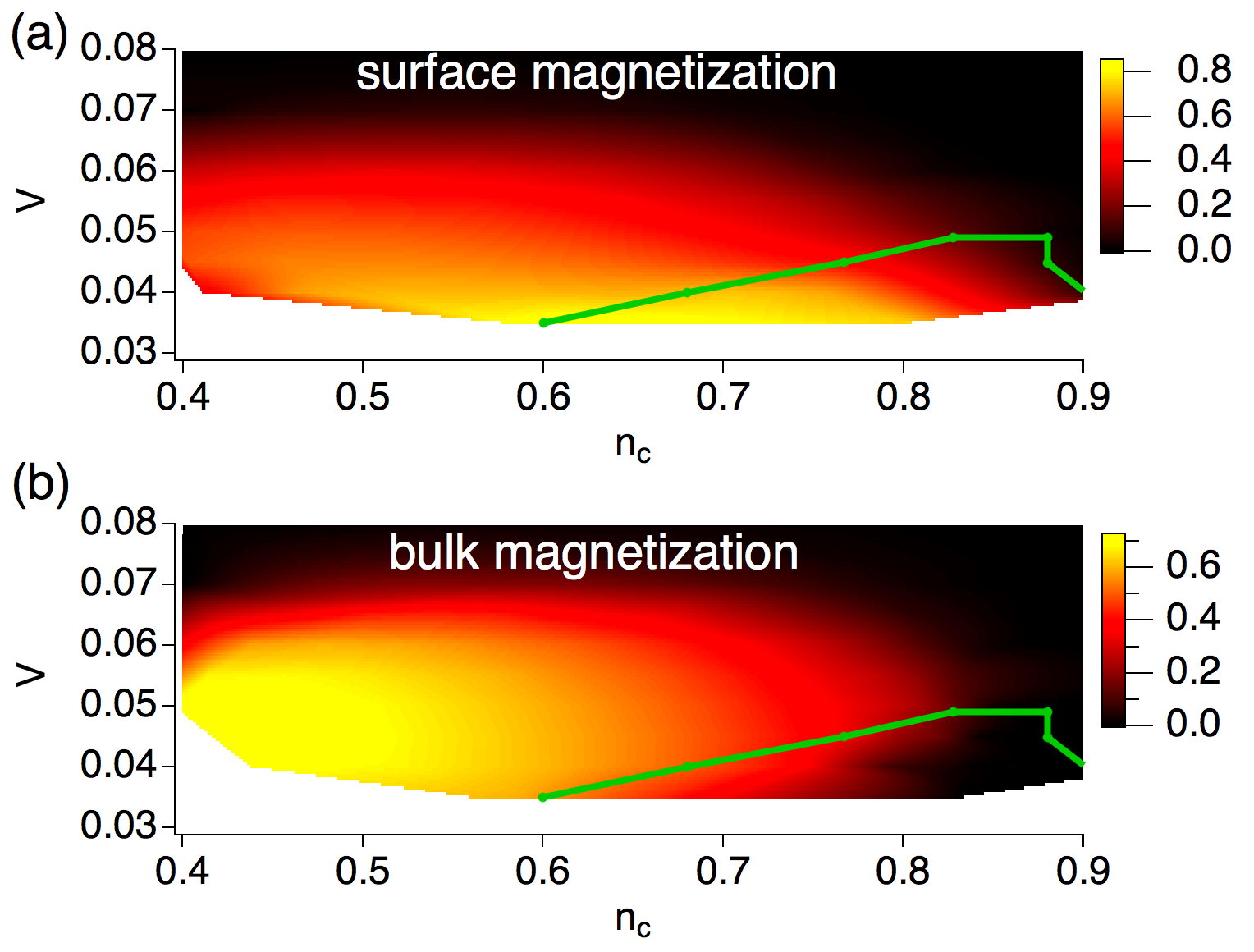}
\end{center}
\caption{Phase diagram for different number of $c$-electrons, $n_c$, and hybridization strengths, $V$, for a system consisting of $20$ layers with open boundaries. A $c$-electron filling of $n_c=0.9$ corresponds to a half-filled system $n_c+n_f=2$. (a) Magnetization for the surface layer. (b) Magnetization of the layer in the middle of the slab. The magnetic phase inside the green line corresponds to a "surface magnetic state" which is A-type antiferromagnetic. The white area at small $V$ in the phase diagram corresponds to a magnetic phase which cannot be described by our ansatz.
\label{Fig2}}
\end{figure}
\begin{figure}[t]
\begin{center}
    \includegraphics[width=0.99\linewidth]{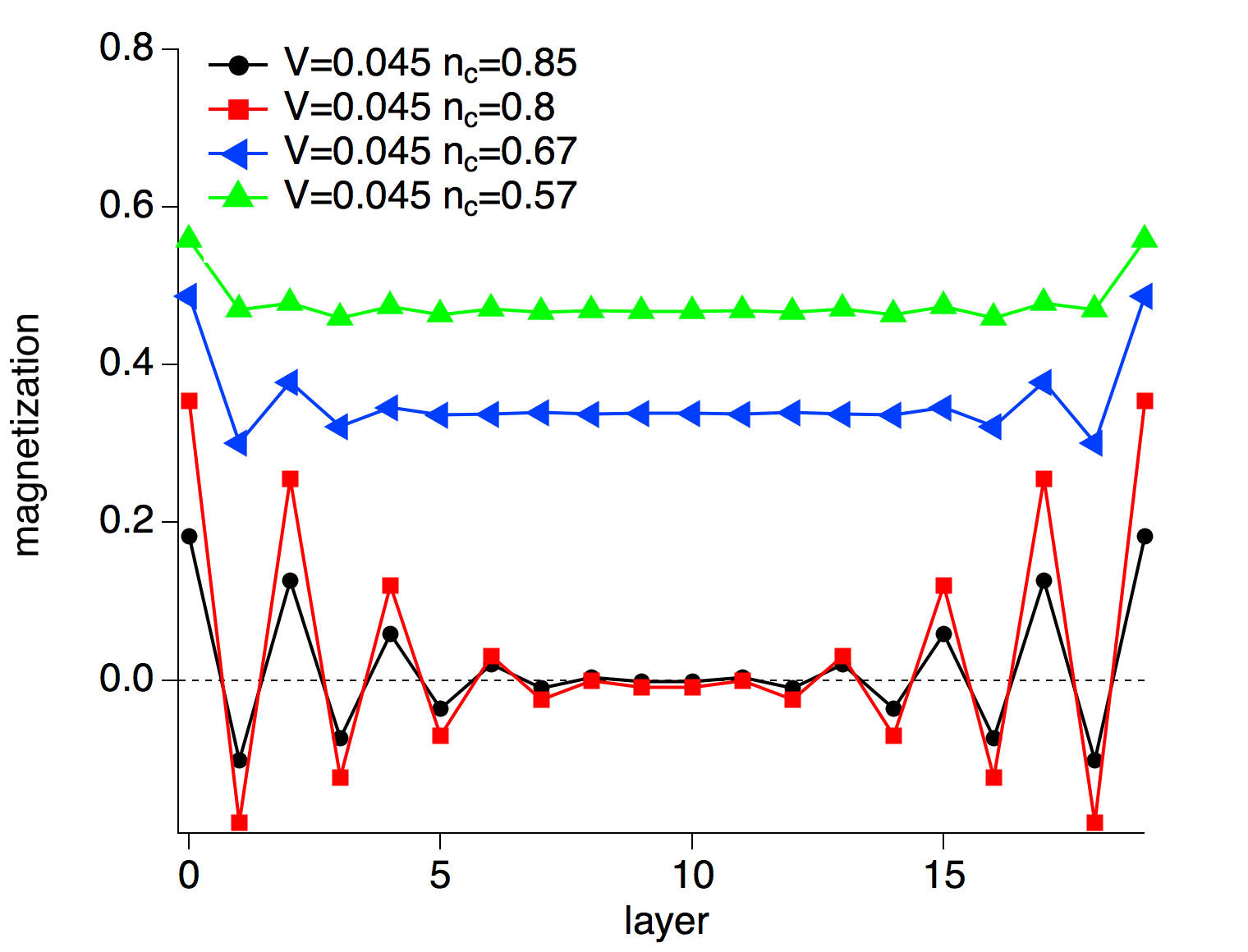}
\end{center}
\caption{Magnetizations for different hybridization strengths, $V$, and the number of $c$-electrons, $n_c$, showing the ferromagnetic phase and the surface magnetic phase.
\label{Fig3}}
\end{figure}
Figure \ref{Fig2} depicts the phase diagram obtained in our calculations. The magnetization of the $f$ electrons, $m_f=n_{\uparrow,f}-n_{\downarrow,f}$ is shown for different hybridization strengths and number of conduction electrons, $n_c=n_{\uparrow,c}+n_{\downarrow c}$.  Because the calculations are done for a system with open surfaces, the magnetization is generally different at the surface (panel a) and the layer in the middle of the slab (panel b).  Depending on the hybridization strength, the phase diagram includes three different phases: When the hybridization strength is large, the system forms a nonmagnetic state. At intermediate hybridization strengths, $0.04\le V\le 0.07$ and conduction electron filling $n_c<0.9$, we find in-plane ferromagnetic states. For hybridization strengths $V<0.04$, our calculations do not converge indicating that the magnetic solution cannot be described by in-plane ferromagnetic states. 

This phase diagram fits into the more general Doniach phase diagram,\cite{doniach77} which describes the competition between the Kondo effect and the RKKY interaction. Large hybridization strengths result in a strong screening by the  Kondo effect and thus the formation of nonmagnetic states. Small hybridization strengths result in a weak Kondo screening so that a magnetic state is formed due to the RKKY interaction. Furthermore, in calculations for the Kondo lattice it was found that for small hybridization strengths a phase transition within the magnetic phase can be observed, which qualitatively agrees with the phase transition found here at $V\approx 0.04$. \cite{PhysRevB.92.075103,PhysRevB.96.115158}

Generally, the surface magnetization is larger than the bulk magnetization, which can be understood as an effective increase of correlations at the surface. 
However, besides the normal ferromagnetic state, where surface and bulk are magnetized in the same direction,  we also find states, where the bulk magnetization vanishes while the surface is magnetically polarized. These solutions can be found for $c$-electron fillings close to half-filling and form a region in the phase diagram which is highlighted by a green line in Fig. \ref{Fig2}. Although these states are in-plane ferromagnetic, the magnetization oscillates depending on the layer and vanishes when going from the surface into the bulk. A so-called A-type antiferromagnetic state, with in-plane ferromagnetic and out-of-plane antiferromagnetic order, is formed.\cite{Chang2017} Characteristic magnetization curves are shown in Fig. \ref{Fig3}. Black and red lines show examples of the magnetization of the A-type surface magnetic states. The magnetization oscillates and vanishes in the bulk. The green and blue line, one the other hand, are examples of the ferromagnetic state. For these solutions, the magnetization slightly decreases when going from the surface into the bulk, but it never becomes zero.  

\section{bulk spectral properties}
\begin{figure}[t]
\begin{center}
    \includegraphics[width=0.99\linewidth]{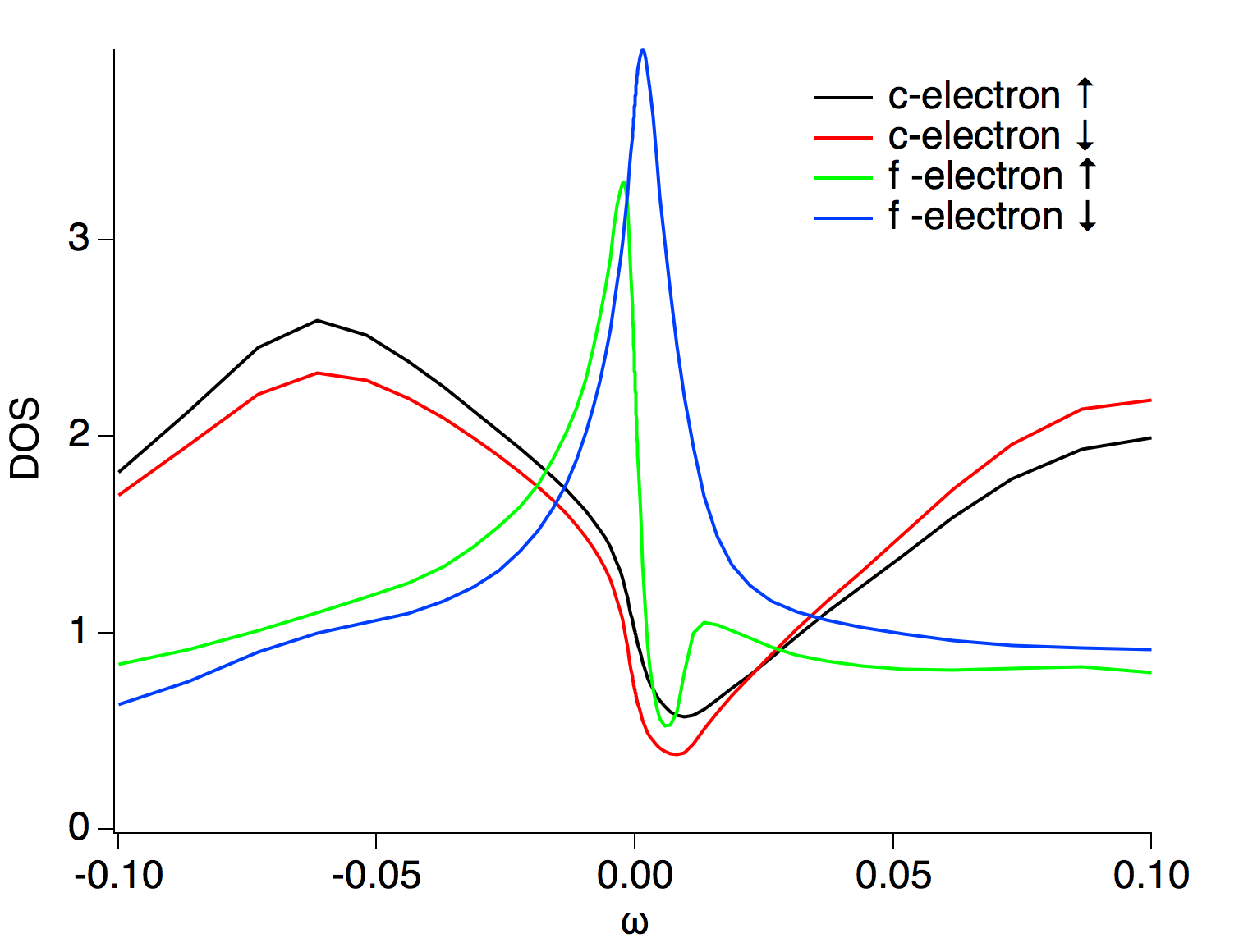}
\end{center}
\caption{Local Spectral functions for the ferromagnetic state $n_c=0.7$ and $V=0.05$  showing separately all four orbitals.  
\label{Fig4}}
\end{figure}
\begin{figure}[t]
\begin{center}
    \includegraphics[width=0.99\linewidth]{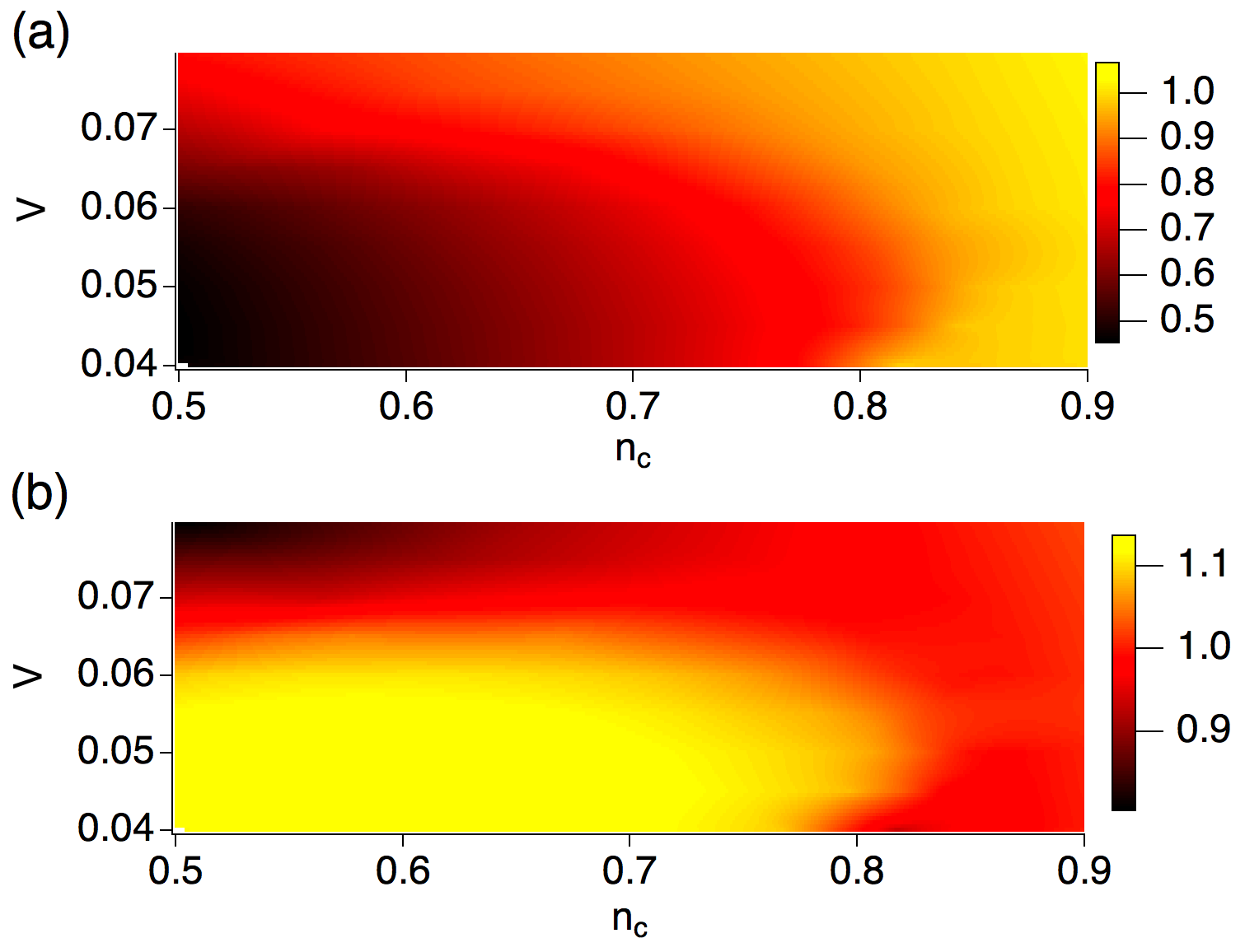}
\end{center}
\caption{Phase diagram as function of the number of $c$-electrons, $n_c$, and hybridization strength, $V$. The plot  shows the particle numbers (a) $n_{c\uparrow}+n_{f\downarrow}$ and  (b) $n_{c\downarrow}+n_{f\uparrow}$ for the middle layer of the slab.
\label{Fig5}}
\end{figure}

Before examining the impact of the magnetic order on the metallic surface states, let us firstly look at the bulk properties. Because states exhibiting surface magnetism are nonmagnetic in the bulk, these states have a renormalized bulk gap similar to nonmagnetic states, which is shifted away from the Fermi energy because of the hole-doping. We will, therefore, focus in this section on states exhibiting bulk ferromagnetism. 

Figure \ref{Fig4} shows a typical local spectral function in the ferromagnetic state for a $c$-electron filling of $n_c=0.7$. The direction of the magnetization is the $z$-direction.

Clearly resolved is a gap structure close to the Fermi energy in the local spectral function for three of the four orbitals and a peak at the Fermi energy for the $f$-electron with down-spin. At first sight, the existence of gap structures close to the Fermi energy might be astonishing, because the total filling of the system is $n_c+n_f=0.7+1.1=1.8$ and thus not half-filled. However, a gap at the Fermi energy for certain spin directions is a commonly observed feature in the ferromagnetic state of a Kondo lattice, resulting in a half-metallic state. \cite{PhysRevB.77.205123,PhysRevB.77.094419,PhysRevB.81.094420,Peters2012,Irkhin1991,PhysRevB.87.165109,PhysRevB.88.054431}
$C$-electrons and $f$-electrons adapt their filling in the ferromagnetic state so that a commensurable situation is created for one of the hybridized spin-sectors. The driving force behind this commensurability condition is the Kondo effect. A closer look at the spectral functions presented in Fig. \ref{Fig4} reveals, however, that the gap is not exactly at the Fermi energy, but slightly above. 

To further investigate this, we directly show the commensurability condition in Fig. \ref{Fig5}. It is important to note that the hybridization used in the model Hamiltonian couples the spins of the $c$- and $f$-electrons using all three Pauli-matrices. Thus, the spin-up (spin-down) component of the $c$-electron is coupled to spin-up and spin-down of the $f$-electron, while in the calculations for a periodic Anderson model showing a perfect commensurability, the spin-up (spin-down) component of the $c$-electron is only coupled to the spin-up (spin-down) component of the $f$-electron. Because there is a coupling between all spin-components, the ferromagnetic state is frustrated. 
In our model Hamiltonian the coupling between the up-spin (down-spin) component of the $c$-electron and down-spin (up-spin) component of the $f$-electron occurs in $x$- and $y$-direction ($\sin k_x \sigma^x$ and $\sin k_y \sigma^y$), while a coupling between the up-spin (down-spin) and the up-spin (down-spin) occurs in $z$-direction ($\sin k_z\sigma^z$). 

We show in Fig. \ref{Fig5} the occupation numbers for $n_{c\uparrow}+n_{f\downarrow}$ (panel a) and $n_{c\downarrow}+n_{f\uparrow}$ (panel b). The bulk ferromagnetic phase is easily visible in panel (b), $n_{c\downarrow}+n_{f\uparrow}$, as an area of constant occupation, although the conduction electron number is varied. On the other hand, in panel (a), $n_{c\uparrow}+n_{f\downarrow}$, there is no area of constant occupation. We can thus identify the spin-down component of the $c$-electron combined with the up-spin component of the $f$-electron as the spin  sector with commensurability. However, because of the frustration occurring due to the hybridization in $z$ direction, the combined occupation is not unity, but slightly larger than one. As a consequence, the gap, which is visible in the local density of states for these orbitals, is slightly shifted above the Fermi energy. 

\begin{figure}[t]
\begin{center}
    \includegraphics[width=0.99\linewidth]{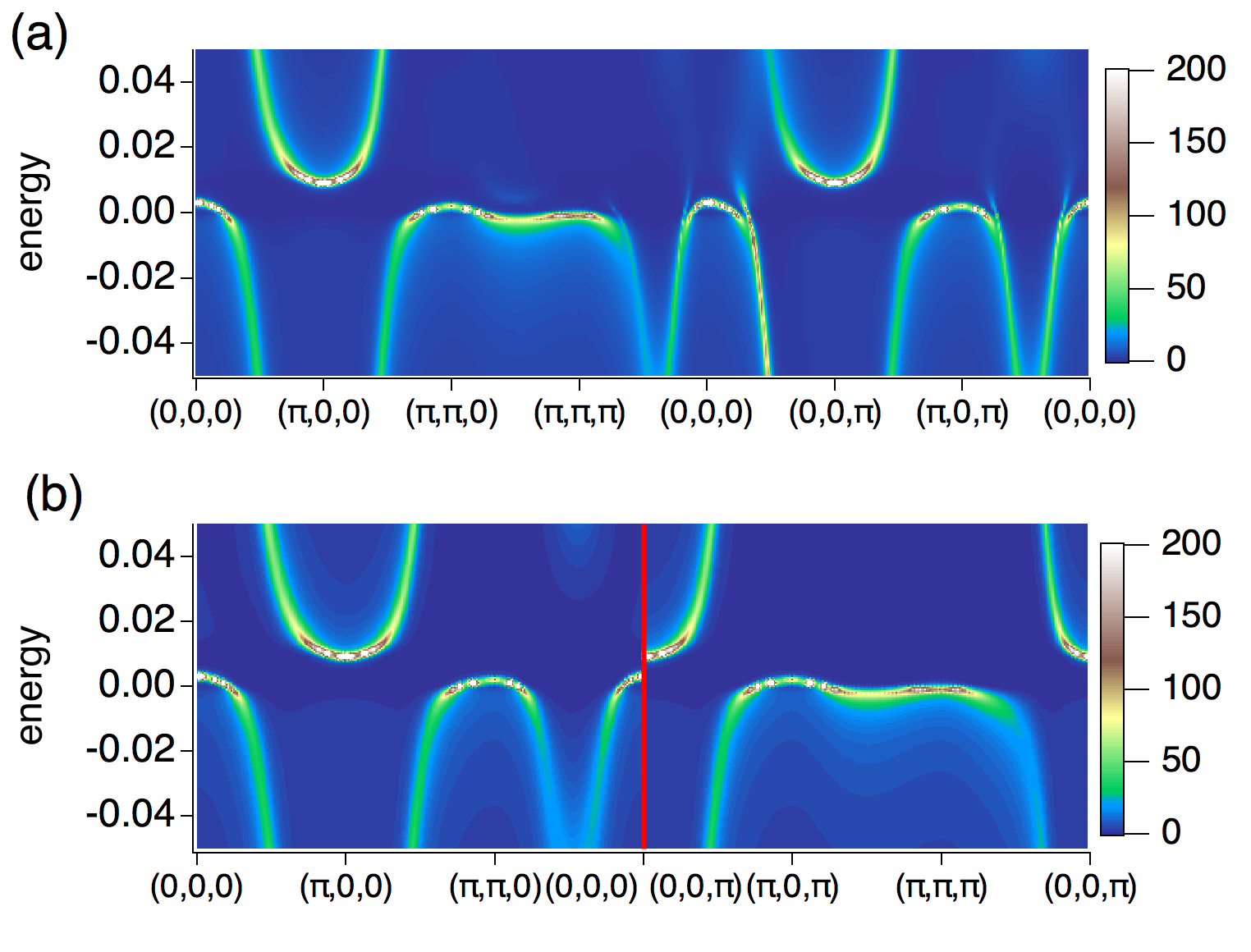}
\end{center}
\caption{Momentum resolved spectral functions of the bulk ferromagnetic state for the spin sector \{$\vert c\downarrow \rangle,\vert f\uparrow \rangle$\} . Panel (a) shows  a cut through the whole Brillouin zone. Panel (b) shows cuts through the $k_x-k_y$-plane for fixed $k_z$, which jumps at the red line from $0$ to $\pi$.
\label{Fig6}}
\end{figure}
\begin{figure}[t]
\begin{center}
    \includegraphics[width=0.99\linewidth]{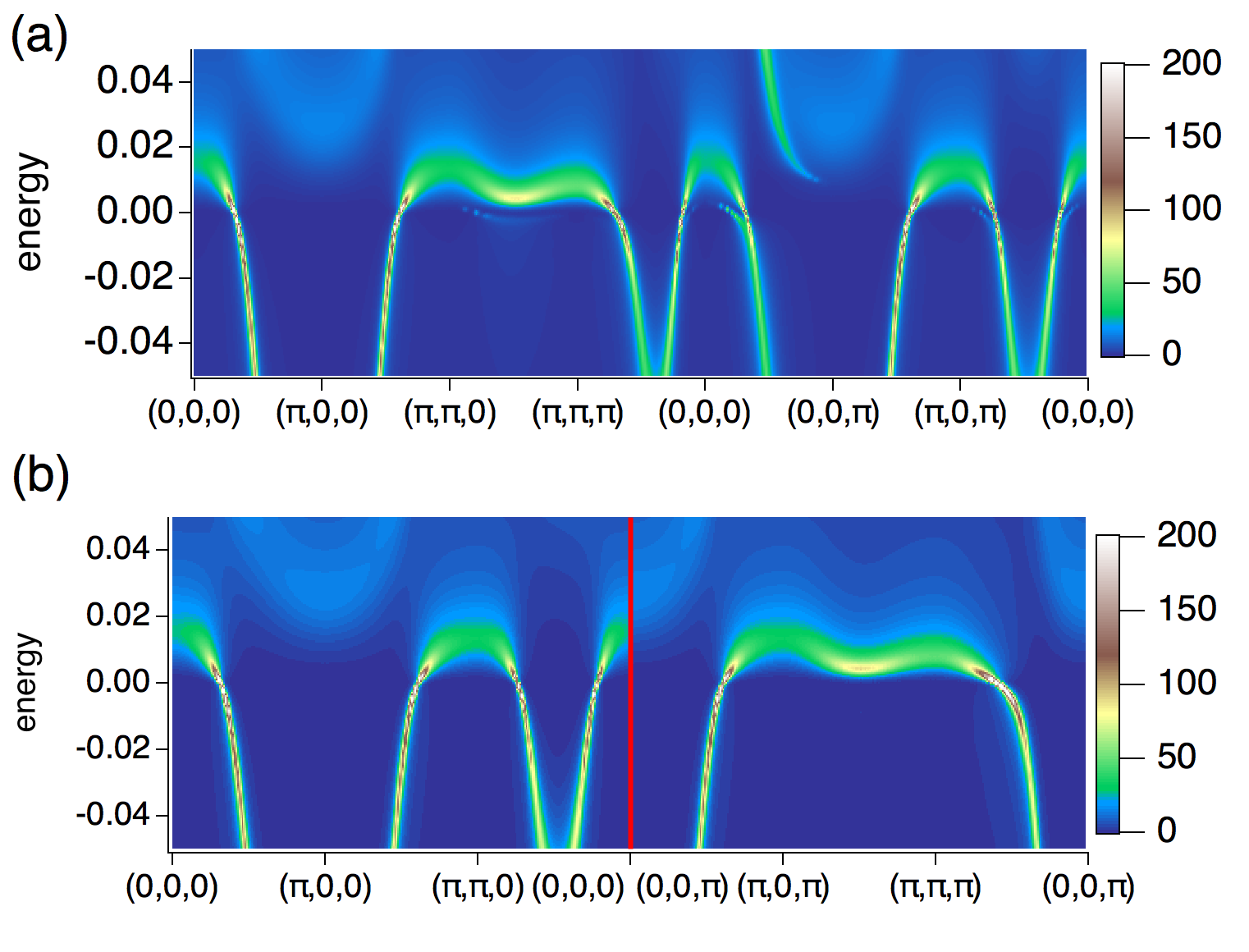}
\end{center}
\caption{Same as Fig. \ref{Fig6}, but for the spin sector \{$\vert c\uparrow \rangle,\vert f\downarrow \rangle$\}.
\label{Fig7}}
\end{figure}
In order to obtain more information about the bulk excitation spectrum, we show momentum-resolved spectral functions of both spin-sectors in Figs. \ref{Fig6} and \ref{Fig7}.
Figure \ref{Fig6} shows the spectral function for the spin-sector \{$\vert c\downarrow \rangle,\vert f\uparrow \rangle$\}, which approximately fulfills the commensurability condition   $n_{c\downarrow}+n_{f\uparrow}\approx1$.
The gap observed in the local spectral functions is also clearly visible here. Looking at Fig. \ref{Fig6}(a), which shows a cut through the whole 3D Brillouin zone, we see that 
bands enter into the gap (see panel (a) between $(\pi,\pi,\pi)\rightarrow(0,0,0)\rightarrow(0,0,\pi)$). Thus, the gap structure visible in Fig. \ref{Fig4} is not a full gap. 
It is however instructive to constrain the momentum space to $k_z=0$ and $k_z=\pi$, whose spectral functions are shown in panel (b). For these momenta, ($k_x,k_y,k_z=0$) and ($k_x,k_y,k_z=\pi$), the coupling between spin-up (spin-down) component of the $c$-electrons and the  spin-up (spin-down) component of the $f$-electrons vanishes. For these momenta, a situation similar to the periodic Anderson model for which a full gap has been observed is reproduced. Indeed, bands do not enter the gap for these momentum planes; for these momenta we find a full gap for the spin sector with approximate commensurability condition. 

The momentum resolved spectral function for the other spin-sector, \{$\vert c\uparrow \rangle,\vert f\downarrow \rangle$\}, is shown in Fig. \ref{Fig7}. A general feature of the spectral function of the spin sector without commensurability is the strong correlation effect, which leads to a strong broadening  around the gap.  Thus, there is no real gap for this spin-sector, but an energy region without quasi-particle bands.
However, a closer look at the spectral function shown in Fig. \ref{Fig7}(a), which shows a cut through the 3D Brillouin zone, reveals that there is at least one quasi-particle band which enters this "gap" region between $(0,0,0)\rightarrow(0,0,\pi)$. If we constrain the plot to the ($k_x,k_y,k_z=0$) and ($k_x,k_y,k_z=\pi$ momentum planes, see Fig. \ref{Fig7}(b), we can see that this band is absent. 

\section{surface states}
Up to now, we have looked at the bulk properties of the system and found a ferromagnetic phase. Next, we want to analyze the effect of the magnetic phase on the symmetry protected surface states, which manifests themselves as Dirac cones in the momentum resolved spectrum. 
\begin{figure}
\begin{center}
    \includegraphics[width=\linewidth]{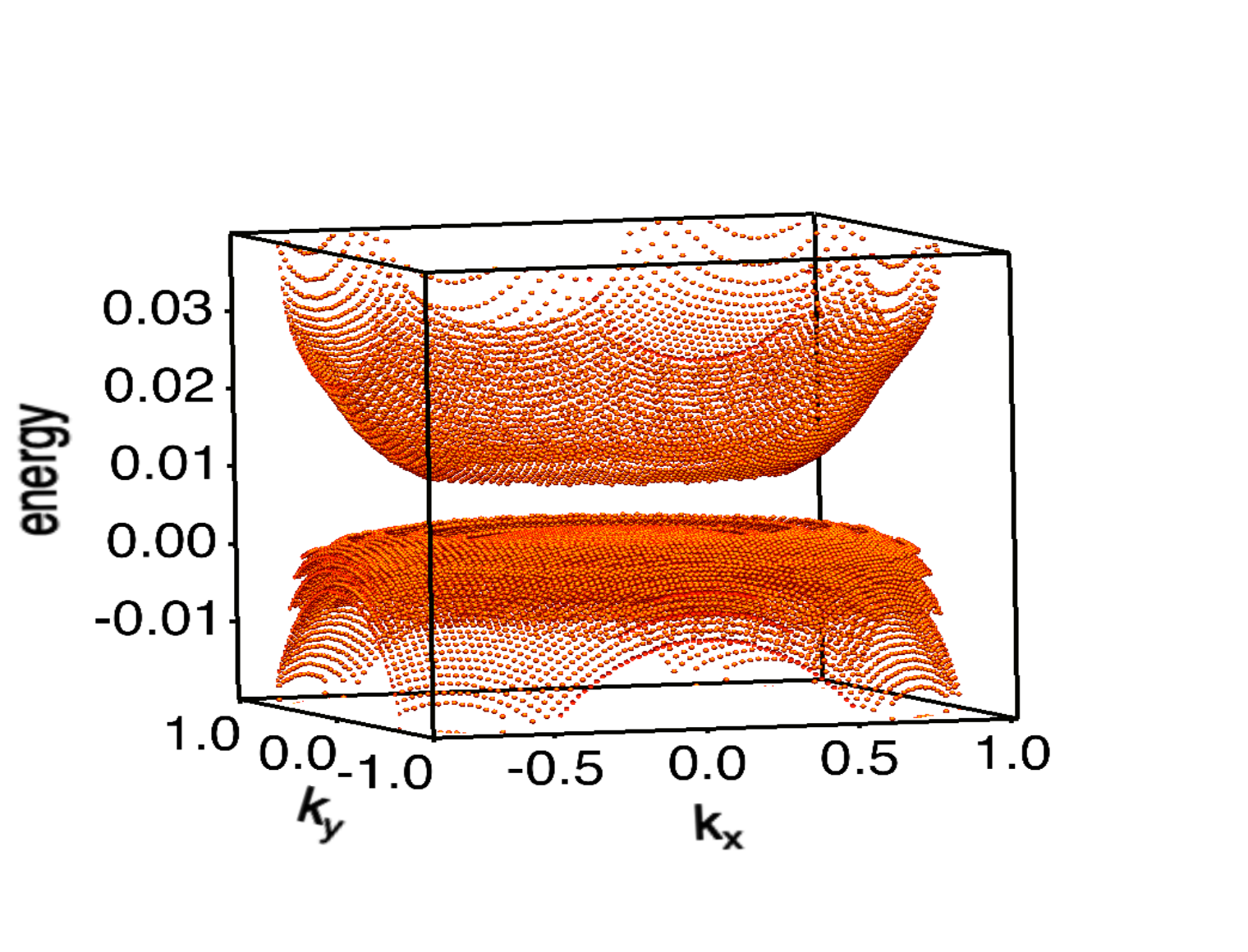}
\end{center}
\caption{Momentum resolved spectrum for $n_c=0.7$, $V=0.05$ in the ferromagnetic phase with magnetization in $z$-direction for the \{$\vert c\downarrow \rangle,\vert f\uparrow \rangle$\} spin sector. The calculation is done for a slab of $20$ layers with open boundary condition in $z$-direction. Thus, the magnetization is perpendicular to the surface. The plot shows the spectrum of the surface layer for $(k_x,k_y)\in([-1,1],[-1,1])$ for which in the noninteracting spectrum a Dirac cone exists. 
\label{Fig8}}
\end{figure}
\begin{figure}
\begin{center}
    \includegraphics[width=\linewidth]{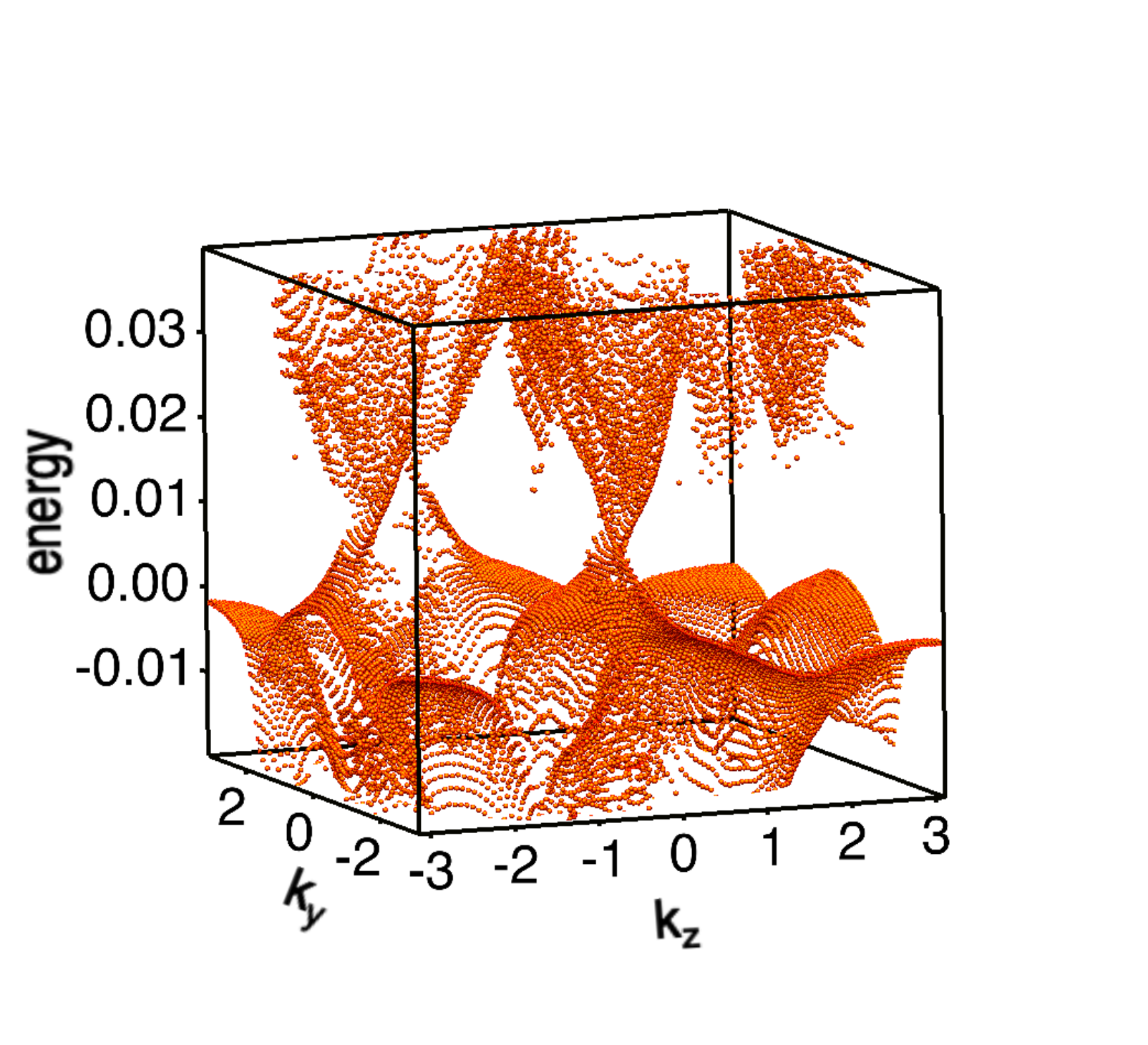}
\end{center}
\caption{The same as Fig. \ref{Fig8}, but with open surface in $x$-direction, which corresponds to an in-plane magnetization at the surface.
\label{Fig9}}
\end{figure}
\begin{figure}
\begin{center}
    \includegraphics[width=\linewidth]{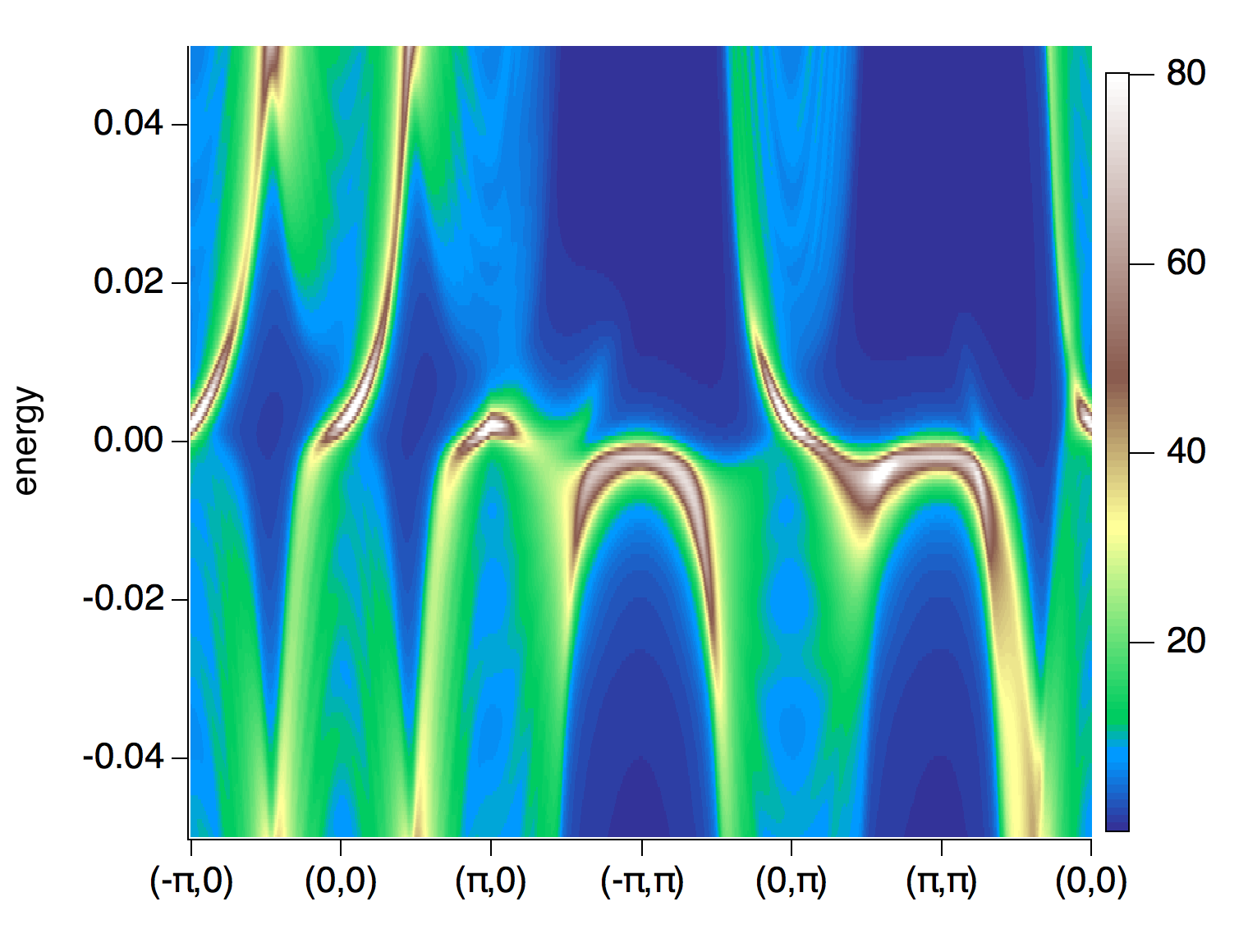}
\end{center}
\caption{Momentum resolved spectrum for $n_c=0.7$, $V=0.05$ in the ferromagnetic phase for the \{$\vert c\downarrow \rangle,\vert f\uparrow \rangle$\} spin sector. The plot shows the spectral intensity on the surface layer in $x$-direction. The momentum is labeled as $(k_y,k_z)$. The momentum $k_y$ increases on the path betweem $(k_y,k_z)=(-\pi,\pi)\rightarrow(0,\pi)\rightarrow(\pi,\pi)$.
\label{Fig10}}
\end{figure}
The Dirac cones at the open surface in the noninteracting spectrum, which are located in the Brillouin zone at $(0,0)$, $(\pi,0)$, and $(0,\pi)$, are protected by time-reversal symmetry. Thus, it is not astonishing to find that in a ferromagnetically ordered system (magnetization in $z$-direction) the surface states have vanished on the $z$-surface. Figure \ref{Fig8} shows the momentum resolved spectral function ($k_x$, $k_y$, energy) of the $z$-surface for $V=0.05$ and $n_c=0.7$. 
Figure \ref{Fig8} was thereby obtained by computing the spectral function for fixed momentum $(k_x,k_y)$. Whenever there is a peak in the spectral function depending on the energy, we plot a single dot. Thus, the spectrum does not include information about the height of the peak in the spectral function. 
Due to the bulk magnetization of $n_{f\uparrow}-n_{f\downarrow}=0.27$ which increases at the surface to $n_{f\uparrow}-n_{f\downarrow}=0.35$ the surface states are fully gapped; there is no Dirac cone visible in the surface spectrum. 

However, the situation is different when looking at different surfaces. While in Fig. \ref{Fig8} we analyze the $z$-surface for a magnetization in $z$-direction, in Fig. \ref{Fig9} we show the surface spectrum in $x$-direction (the magnetization is still in $z$-direction)  for the \{$\vert c\downarrow \rangle,\vert f\uparrow \rangle$\}  spin sector. Thus, this situation corresponds to an in-plane magnetic state. We see at the first sight that the spectrum is not gapped. Taking into account the knowledge about the bulk spectrum, we conclude that these bands are surface states. In Fig. \ref{Fig10}, we show the momentum resolved spectral function of the surface layer for a cut through the surface Brillouin zone, which also clearly shows states going through the bulk gap. The position of these surface states is thereby approximately at the same momenta as the symmetry protected Dirac cones  in the nonmagnetic system, namely at $(k_y,k_z)=(0,0)$, $(\pi,0)$, and $(0,\pi)$. 

How can we understand the existence of these surface states and are these surface states protected by any symmetries or just accidental?  As the time-reversal symmetry is broken by the magnetization, this symmetry cannot protect any surface states spanning the gap. The answer to this question comes here from the cubic symmetry of the model Hamiltonian, a symmetry which is also preserved in the Kondo insulator SmB$_6$. 
Because of the cubic symmetry, the Hamiltonian 
conserves the following reflection symmetry, even in the presence of a magnetization along the $z$-direction: $R_z=i\sigma^z \tau^z P_z$,
where $\sigma^z$ and $\tau^z$ are Pauli matrices  acting on the spin-indices and orbital-indices, respectively; $P_z$ flips the sign of $k_z$ ($P_z:\quad k_z\rightarrow -k_z$).  
In the case of a magnetization in $x$- or $y$-direction we can define operators $R_x=i\sigma^x\tau^zP_x$ or $R_y=i\sigma^y\tau^zP_y$, which still commute with the Hamiltonian.

The presence of this symmetry guarantees that the Hamiltonian can be separated into two subspaces, which do not couple to each other even in the presence of a magnetization.  
In the case of a magnetization in $z$-direction, the Hamiltonian is separated on the reflection invariant planes in the Brillouin zone, $k_z=0$ or $k_z=\pi$, into the  \{$\vert c\uparrow \rangle,\vert f\downarrow \rangle$\} spin sector and  \{$\vert c\downarrow \rangle,\vert f\uparrow \rangle$\} spin sector which correspond to the plus- ($R_z\Psi=\Psi$) and minus- ($R_z\Psi=-\Psi$) subspaces of the reflection operator, respectively. 
Thus, this separation shown above is not accidental, but originates in the reflection symmetry. \cite{PhysRevLett.111.226403,PhysRevLett.115.156405,PhysRevLett.106.106802,PhysRevLett.93.206602}

From now on, we will focus on the minus sector of the reflection operator, \{$\vert c\downarrow \rangle,\vert f\uparrow \rangle$\} spin sector, constrained to the $k_z=0$ or $k_z=\pi$ plane of the Brillouin zone. We have demonstrated above that this sector is gapped in the bulk when constrained to these planes in the Brillouin zone. Thus, the Chern number is well defined. 
In the presence of  electron correlations, the Chern number can be calculated from the Green's function as\cite{PhysRevB.83.085426,PhysRevB.85.125113,ISHIKAWA1987523}
\begin{equation}
N=\frac{\epsilon^{\mu\nu\rho}}{24\pi^2}\int d^3k \mathrm{Tr}\left( G^{-1}\frac{\partial  G}{\partial k_\mu} G^{-1}\frac{\partial  G}{\partial k_\nu}  G^{-1}\frac{\partial  G}{\partial k_\rho}  \right),
\end{equation}
with $k:=(\omega,k_x,k_y)$. $\epsilon^{\mu\nu\rho}$ denotes the total anti-symmetric Levi-Civita symbol satisfying $\epsilon^{012}=1$. 
The Green's function is defined on the imaginary axis $G(k):=G(i\omega,\bm{k})$.

In the work of \textcite{PhysRevX.2.031008}, it was shown that as long as the self-energy is non-singular, replacing the full Green's function $G$ with the simplified Green's function
\begin{eqnarray}
\tilde G^{-1}(i\omega,\bm{k})&=& i\omega \rho_0 -H_{eff}(\bm{k}), \\
H_{eff}(\bm{k}) &=&H(\bm{k})+\mathrm{Re}\Sigma(i\omega=0)\label{Heff}, \nonumber \\
                &=&n_0(\bm{k})\rho_0 +\bm{n}(\bm{k})\rho_i,\label{simpGreen}
\end{eqnarray}
does not change the value of the Chern number.
The Pauli matrices $\rho_i$ act on the two states spanning the minus sector of the reflection, \{$\vert c\downarrow \rangle,\vert f\uparrow \rangle$\}.(We have confirmed  the absence of any singularity in the self-energy by direct computation; the imaginary part of the self-energy vanishes around the gap.)
In the case, where the effective Hamiltonian $H_{eff}$ is two-dimensional, a further simplification is possible.
The coefficient-vector $\bm{n}$, which is defined in Eq. (\ref{simpGreen}), can  be used to efficiently calculate the Chern number of the minus sector of the reflection reading
\begin{equation}
N=\frac{1}{4\pi}\int d^2\bm{k} \hat{\vec n}\cdot\left(\frac{\partial \hat{\vec n}}{\partial k_x}\times\frac{\partial \hat{\vec n}}{\partial k_y}\right),
\end{equation}
where $\hat{\vec n}:={\vec n}/\sqrt{ {\vec n} \cdot {\vec n}}$.

Calculating the Chern number for the ferromagnetic phase, we find that the Chern number $N=2$ for $k_z=0$ and $N=-1$ for $k_z=\pi$.
These nonzero Chern numbers are the evidence for the existence of two chiral surface states for $k_z=0$ and one chiral surface state for $k_z=\pi$ spanning the gap in the \{$\vert c\downarrow \rangle,\vert f\uparrow \rangle$\} spin sector, if $R_z=i\sigma^z \tau^z P_z$ is conserved. This means that we have symmetry protected surface states on the surfaces in $x$- and $y$-direction for a magnetization in $z$-direction. Thus, the system realizes a ferromagnetic topological crystalline half-metallic state.
These values of the Chern number can also be easily verified in the spectrum shown in Fig. \ref{Fig10}. For the $k_z=0$ plane, we find two chiral states at $(k_y,k_z)=(-\pi,0)$ and $(0,0)$ running from left to right for increasing energy, which corresponds to the Chern number $N=2$, and one chiral state at $(k_y,k_z)=(0,\pi)$ running from right to left, which corresponds to $N=-1$.

\begin{figure}[t]
\begin{center}
    \includegraphics[width=0.99\linewidth]{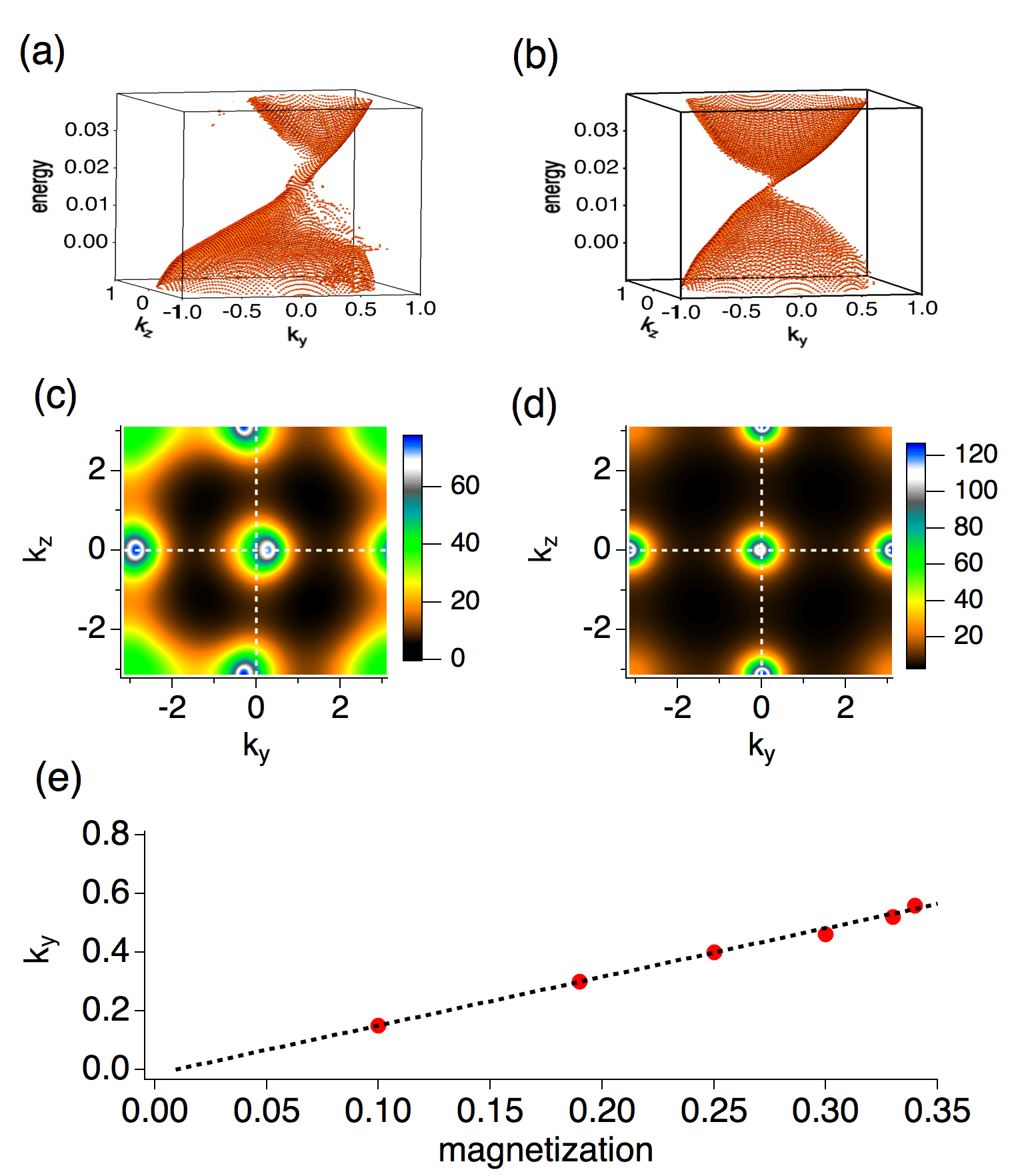}
\end{center}
\caption{Comparison of the surface states between the nonmagnetic system and the ferromagnetic system. (a) Surface state for momenta around $(k_y,k_z)=(0,0)$ in the magnetic state ($V=0.06$, $n_c=0.7$ and $n_{f\uparrow}-n_{f\downarrow}=0.2$). (b)  Same as (a) but for the nonmagnetic state ($V=0.06$, $n_c=0.9$ and $n_{f\uparrow}-n_{f\downarrow}=0.0$). Panels (c) and (d) show cuts through the Brillouin zone for energy $E=0.019$ of panels (a) and (b) respectively. (e) Position of the Dirac cone depending on the surface magnetization for fixed hybridization strength $V=0.06$ changing the $c$-electron filling.
\label{Fig11}}
\end{figure}
\begin{figure}[t]
\begin{center}
    \includegraphics[width=0.99\linewidth]{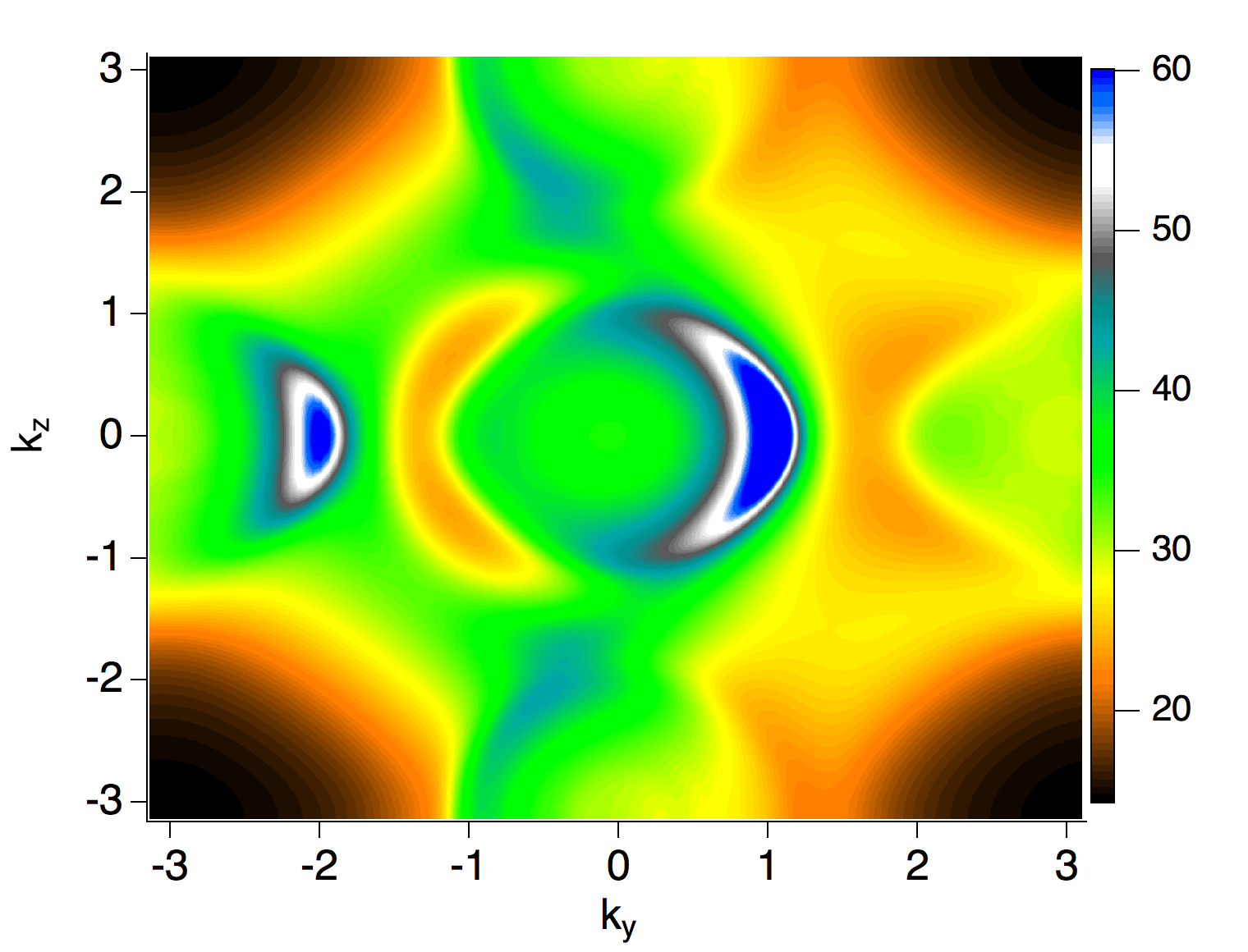}
\end{center}
\caption{Momentum resolved spectral function of the surface layer for $n_c=0.5$ and $V=0.04$ with surface magnetization $n_{f\uparrow}-n_{f\downarrow}=0.6$. The energy of the spectral function is fixed to $E=0.005$ below the center of the Dirac cone. Due to the magnetization, the Dirac cone appears not as a ring in the spectral function, but as an arc.
\label{Fig12}}
\end{figure}
We have seen in the paragraph above that the existence of surface states on the $x$-surface for a ferromagnetic state with magnetization in $z$-direction is protected by symmetry in our Hamiltonian. We next want to ask, what is the impact of the magnetization on the surface states, which resemble Dirac cones in the nonmagnetic system? The symmetry protection due to nontrivial topology works in the ferromagnetic system only for the reflection invariant planes; $k_z=0$ and $k_z=\pi$ for a magnetization in $z$-direction. The topological surface states in the ferromagnetic system  constrained to these momentum planes take the form of chiral edge modes. 
Away from these planes the symmetry protection due to the reflection symmetry does not work.  In the nonmagnetic system, on the other hand, we have full rotational invariant Dirac cones which are formed by states from both spin-sectors, \{$\vert c\uparrow \rangle,\vert f\downarrow \rangle$\} and \{$\vert c\downarrow \rangle,\vert f\uparrow \rangle$\}.

Figure \ref{Fig11}(a)-(d) show a comparison between the surface state of the nonmagnetic system ($n_c=0.9$, $V=0.06$) and the ferromagnetic state ($n_c=0.7$, $V=0.06$).  Figures \ref{Fig11}(a) and  \ref{Fig11}(b) show three dimensional plots ($k_y$-$k_z$-energy) of the surface state on the surface in $x$-direction for the magnetic and nonmagnetic system, respectively.  Figures \ref{Fig11}(c) and  \ref{Fig11}(d) show intensity plots for fixed energy $E=0.019$, approximately at the center of the Dirac cone in the nonmagnetic system.

We immediately see from the comparison in Fig. \ref{Fig11} (a) and (b) that the Dirac cone is strongly deformed by the ferromagnetic state. While the lower dome of the Dirac cone (energy $E<0.02$) remains approximately as it is, the upper dome and the Dirac point (energy $E\ge 0.02$) are moved to different momenta. For a better comparison we show intensity plots of the spectral function for $E=0.019$ in panels (c) and (d). Approximately at this energy the Dirac cone contracts to a single point in the Brillouin zone.  The spectral function shown here is the sum of all orbitals calculated at the surface layer. 
Figure \ref{Fig11}(d) shows the nonmagnetic state. Clearly visible are regions of high intensity (white-blue) in the spectral function around $(k_y,k_z)=(0,0)$, $(\pi,0)$, $(0,\pi)$. These regions of high intensity correspond to the Dirac cones in the spectrum and are located at the high symmetry points in the Brillouin zone.
Figure \ref{Fig11}(c) shows the spectral function at the same energy for the ferromagnetic state of panel (a). 
It is clearly seen that the high intensity regions in the spectral function have shifted away from the high symmetry points in the Brillouin zone. The Dirac cones on the $k_z=0$ plane is shifted to the right and the Dirac cone on the $k_z=\pi$ plane is shifted to the left. The Dirac cones are thereby still perfectly located on the planes with $k_z=0$ and $k_z=\pi$, which is due to the reflection symmetry. Because we add the spectral intensity of all orbitals in these plots, there is also a rather high intensity at $(k_y,k_z)=(\pi,\pi)$. This density of states originates in the \{$\vert c\uparrow \rangle,\vert f\downarrow \rangle$\}- spin sector which is not fully gapped. In Fig. \ref{Fig11}(e), we finally show the position of the Dirac cone depending on the surface magnetization. The position of the Dirac cone is calculated by finding the maximum of the spectral density for $k_z=0$ and $E=0.019$ depending on $k_y$. We see that the position of the Dirac cone behaves linearly with the surface magnetization.

Above we have seen that the Dirac cones in the nonmagnetic system are changed to chiral states protected by reflection symmetry in the ferromagnetic system and seem to be shifted away
 from the high symmetry points of the Brillouin zone. However, this is not the only effect on the surface states. The Dirac cones in the nonmagnetic system consists of $c$-electrons and $f$-electrons with up- and down-spin direction. On the ferromagnetically polarized surface, electrons with different spin-direction have different energy. The consequence of this is shown in Fig. \ref{Fig12}, which shows the surface spectrum for energy $E=0.005$, which cuts through the Dirac cone. The spectrum includes all bands and spin-directions. In the case of the nonmagnetic system, the spectrum shows rings of high intensity around the high symmetry points where the Dirac cones are located. In Fig. \ref{Fig12}, we show that in the ferromagnetic system an arc instead of the ring is observed. The origin of this arc in the spectrum is different from the Weyl-semimetal. It arises because of the magnetic polarization of the Dirac cone. There is only one half of the Dirac cone present.The other half of the Dirac cone has vanished due to the energy shift of the electrons. 
\begin{figure}[t]
\begin{center}
    \includegraphics[width=0.99\linewidth]{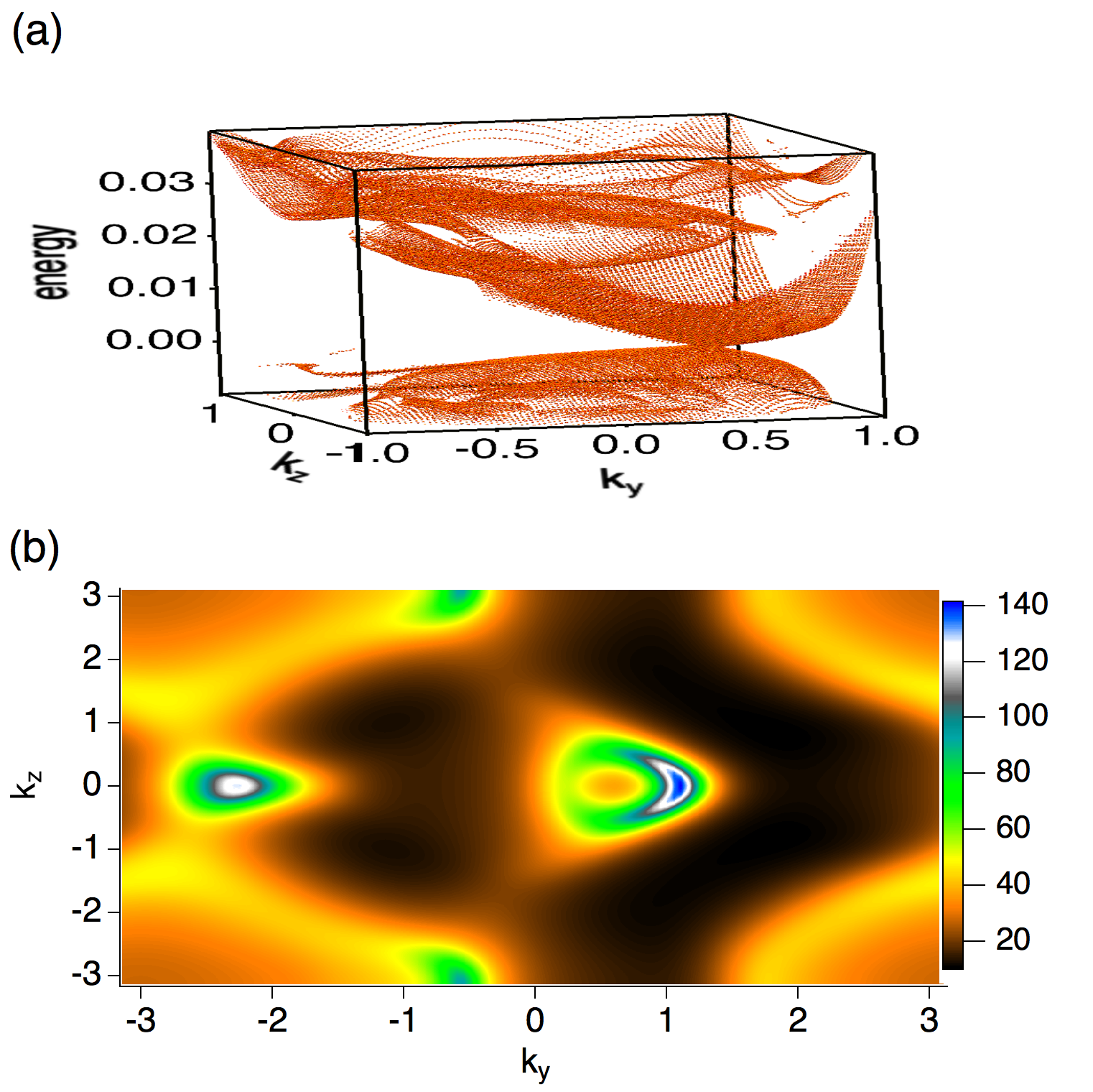}
\end{center}
\caption{Spectrum of the A-type antiferromagnetic state. (a) ($k_y$-$k_z$-energy) plot of the spectrum on an $x$-surface. (b) Momentum-resolved spectrum for $E=0.005$.
\label{Fig13}} 
\end{figure}

Before concluding, let us present some results about the  A-type antiferromagnetic state, where the magnetization vanishes in the bulk. First, because the bulk is nonmagnetic, time-reversal symmetry and reflection symmetry are conserved in the bulk. Thus, the symmetry protection of topological states holds in the bulk. 
At the surface, the time-reversal symmetry is broken, but the reflection symmetry for the direction of the magnetization is conserved. A similar protection as in the ferromagnetic state could work. On the other hand, due to the antiferromagnetic ordering of different layers, the spin-sectors \{$\vert c\uparrow \rangle,\vert f\downarrow \rangle$\} and \{$\vert c\downarrow \rangle,\vert f\uparrow \rangle$\} are not separated any more. Thus, the surface states could be gapped out due to hybridization with the other spin-sector. However, our results show that even for the antiferromagnetic surface state, Dirac-type surface states still exist, see Fig. \ref{Fig13}. Because of the mixing of different spin-sectors, a gapped sector does not exist anymore, the spectrum at the surface includes different bands. Focusing on the momenta around $(k_y,k_z)=(0,0)$, we see a deformed Dirac cone which is shifted from $(k_y,k_z)=(0,0)$ to approximately $(k_y,k_z)=(1,0)$ due to a strong surface magnetization, see Fig. \ref{Fig13}(a). The spectral intensity for energy $E=0.005$ is shown in Fig. \ref{Fig13}(b). We observe two regions of high intensity (green-blue) for $k_z=0$ and one region for $k_z=\pi$, which coincides with the existence of the Dirac cones in the ferromagnetic state. Thus, we conclude that even for this antiferromagnetic state the symmetry protected surface states exist. Furthermore, we see that the Dirac cone at $(k_y,k_z)=(1,0)$ has an arc shape due to the magnetic polarization. Besides these deformed Dirac cones,  we see broad bands with intermediate intensity (orange) around  $(k_y,k_z)=(\pi,\pi)$. 

\section{Conclusions}
We have analyzed the possibility of magnetically ordered states in a 3D cubic topological Kondo insulator. We have demonstrated the existence of a wide ferromagnetic phase which emerges upon hole doping. Besides this phase, we find surface magnetic states close to half-filling, which are A-type antiferromagnetically ordered. 
While in the nonmagnetic system there are symmetry protected surface states on all surfaces of this system, the surface states are gapped out in the ferromagnetic state when the magnetization is perpendicular to the surface. Surface states for layers with in-plane magnetization are thereby protected by reflection symmetry in our model, which is also conserved in the candidate topological Kondo insulator SmB$_6$. The emergence or absence of surface states depending on the magnetization direction could thereby yield interesting technological applications. Switching the magnetization direction by an external magnetic field would generate or destroy the surface states spanning the gap. 

We have furthermore elucidated the impact of the magnetization on the surface states, which manifests themselves as Dirac cones in the nonmagnetic system. The magnetization shifts the Dirac cones away from the high symmetry points in the surface Brillouin zone. The shift is thereby proportional to the surface magnetization. Furthermore, while in the nonmagnetic system Dirac cones appear as rings in the momentum resolved spectrum at fixed energy, these surface states are deformed into arcs due to the magnetization. The arc thereby occurs due to the energy shift of certain spin-directions.

This study shows that the interplay between strong correlations and nontrivial topology has quite a few of novel phenomena to be explored, which might be also used   in future applications.

\begin{acknowledgments}
This work is partly supported by JSPS KAKENHI Grant No. 25220711, JP15H05855,  JP16K05501, 18K03511, and No. 18H04316 and CREST, JST No. JPMJCR1673.
Computer simulations were performed on the "Hokusai" supercomputer in RIKEN and the supercomputer of  the Institute for Solid State Physics (ISSP) in Japan.  \end{acknowledgments}
\bibliography{paper}

\begin{thebibliography}{58}
\expandafter\ifx\csname natexlab\endcsname\relax\def\natexlab#1{#1}\fi
\expandafter\ifx\csname bibnamefont\endcsname\relax
  \def\bibnamefont#1{#1}\fi
\expandafter\ifx\csname bibfnamefont\endcsname\relax
  \def\bibfnamefont#1{#1}\fi
\expandafter\ifx\csname citenamefont\endcsname\relax
  \def\citenamefont#1{#1}\fi
\expandafter\ifx\csname url\endcsname\relax
  \def\url#1{\texttt{#1}}\fi
\expandafter\ifx\csname urlprefix\endcsname\relax\def\urlprefix{URL }\fi
\providecommand{\bibinfo}[2]{#2}
\providecommand{\eprint}[2][]{\url{#2}}

\bibitem[{\citenamefont{Hasan and Kane}(2010)}]{RevModPhys.82.3045}
\bibinfo{author}{\bibfnamefont{M.~Z.} \bibnamefont{Hasan}} \bibnamefont{and}
  \bibinfo{author}{\bibfnamefont{C.~L.} \bibnamefont{Kane}},
  \bibinfo{journal}{Rev. Mod. Phys.} \textbf{\bibinfo{volume}{82}},
  \bibinfo{pages}{3045} (\bibinfo{year}{2010}),
  \urlprefix\url{http://link.aps.org/doi/10.1103/RevModPhys.82.3045}.

\bibitem[{\citenamefont{Qi and Zhang}(2011)}]{RevModPhys.83.1057}
\bibinfo{author}{\bibfnamefont{X.-L.} \bibnamefont{Qi}} \bibnamefont{and}
  \bibinfo{author}{\bibfnamefont{S.-C.} \bibnamefont{Zhang}},
  \bibinfo{journal}{Rev. Mod. Phys.} \textbf{\bibinfo{volume}{83}},
  \bibinfo{pages}{1057} (\bibinfo{year}{2011}),
  \urlprefix\url{http://link.aps.org/doi/10.1103/RevModPhys.83.1057}.

\bibitem[{\citenamefont{Pesin and Balents}(2010)}]{Pesin2010}
\bibinfo{author}{\bibfnamefont{D.}~\bibnamefont{Pesin}} \bibnamefont{and}
  \bibinfo{author}{\bibfnamefont{L.}~\bibnamefont{Balents}},
  \bibinfo{journal}{Nat Phys} \textbf{\bibinfo{volume}{6}},
  \bibinfo{pages}{376} (\bibinfo{year}{2010}), ISSN \bibinfo{issn}{1745-2473},
  \urlprefix\url{http://dx.doi.org/10.1038/nphys1606}.

\bibitem[{\citenamefont{Hohenadler et~al.}(2011)\citenamefont{Hohenadler, Lang,
  and Assaad}}]{PhysRevLett.106.100403}
\bibinfo{author}{\bibfnamefont{M.}~\bibnamefont{Hohenadler}},
  \bibinfo{author}{\bibfnamefont{T.~C.} \bibnamefont{Lang}}, \bibnamefont{and}
  \bibinfo{author}{\bibfnamefont{F.~F.} \bibnamefont{Assaad}},
  \bibinfo{journal}{Phys. Rev. Lett.} \textbf{\bibinfo{volume}{106}},
  \bibinfo{pages}{100403} (\bibinfo{year}{2011}),
  \urlprefix\url{http://link.aps.org/doi/10.1103/PhysRevLett.106.100403}.

\bibitem[{\citenamefont{Yu et~al.}(2011)\citenamefont{Yu, Xie, and
  Li}}]{PhysRevLett.107.010401}
\bibinfo{author}{\bibfnamefont{S.-L.} \bibnamefont{Yu}},
  \bibinfo{author}{\bibfnamefont{X.~C.} \bibnamefont{Xie}}, \bibnamefont{and}
  \bibinfo{author}{\bibfnamefont{J.-X.} \bibnamefont{Li}},
  \bibinfo{journal}{Phys. Rev. Lett.} \textbf{\bibinfo{volume}{107}},
  \bibinfo{pages}{010401} (\bibinfo{year}{2011}),
  \urlprefix\url{http://link.aps.org/doi/10.1103/PhysRevLett.107.010401}.

\bibitem[{\citenamefont{Hohenadler and Assaad}(2013)}]{0953-8984-25-14-143201}
\bibinfo{author}{\bibfnamefont{M.}~\bibnamefont{Hohenadler}} \bibnamefont{and}
  \bibinfo{author}{\bibfnamefont{F.~F.} \bibnamefont{Assaad}},
  \bibinfo{journal}{Journal of Physics: Condensed Matter}
  \textbf{\bibinfo{volume}{25}}, \bibinfo{pages}{143201}
  (\bibinfo{year}{2013}),
  \urlprefix\url{http://stacks.iop.org/0953-8984/25/i=14/a=143201}.

\bibitem[{\citenamefont{Yoshida et~al.}(2012)\citenamefont{Yoshida, Fujimoto,
  and Kawakami}}]{PhysRevB.85.125113}
\bibinfo{author}{\bibfnamefont{T.}~\bibnamefont{Yoshida}},
  \bibinfo{author}{\bibfnamefont{S.}~\bibnamefont{Fujimoto}}, \bibnamefont{and}
  \bibinfo{author}{\bibfnamefont{N.}~\bibnamefont{Kawakami}},
  \bibinfo{journal}{Phys. Rev. B} \textbf{\bibinfo{volume}{85}},
  \bibinfo{pages}{125113} (\bibinfo{year}{2012}),
  \urlprefix\url{http://link.aps.org/doi/10.1103/PhysRevB.85.125113}.

\bibitem[{\citenamefont{Yoshida et~al.}(2014)\citenamefont{Yoshida, Peters,
  Fujimoto, and Kawakami}}]{PhysRevLett.112.196404}
\bibinfo{author}{\bibfnamefont{T.}~\bibnamefont{Yoshida}},
  \bibinfo{author}{\bibfnamefont{R.}~\bibnamefont{Peters}},
  \bibinfo{author}{\bibfnamefont{S.}~\bibnamefont{Fujimoto}}, \bibnamefont{and}
  \bibinfo{author}{\bibfnamefont{N.}~\bibnamefont{Kawakami}},
  \bibinfo{journal}{Phys. Rev. Lett.} \textbf{\bibinfo{volume}{112}},
  \bibinfo{pages}{196404} (\bibinfo{year}{2014}),
  \urlprefix\url{https://link.aps.org/doi/10.1103/PhysRevLett.112.196404}.

\bibitem[{\citenamefont{Tada et~al.}(2012)\citenamefont{Tada, Peters, Oshikawa,
  Koga, Kawakami, and Fujimoto}}]{PhysRevB.85.165138}
\bibinfo{author}{\bibfnamefont{Y.}~\bibnamefont{Tada}},
  \bibinfo{author}{\bibfnamefont{R.}~\bibnamefont{Peters}},
  \bibinfo{author}{\bibfnamefont{M.}~\bibnamefont{Oshikawa}},
  \bibinfo{author}{\bibfnamefont{A.}~\bibnamefont{Koga}},
  \bibinfo{author}{\bibfnamefont{N.}~\bibnamefont{Kawakami}}, \bibnamefont{and}
  \bibinfo{author}{\bibfnamefont{S.}~\bibnamefont{Fujimoto}},
  \bibinfo{journal}{Phys. Rev. B} \textbf{\bibinfo{volume}{85}},
  \bibinfo{pages}{165138} (\bibinfo{year}{2012}),
  \urlprefix\url{http://link.aps.org/doi/10.1103/PhysRevB.85.165138}.

\bibitem[{\citenamefont{Fidkowski and Kitaev}(2010)}]{PhysRevB.81.134509}
\bibinfo{author}{\bibfnamefont{L.}~\bibnamefont{Fidkowski}} \bibnamefont{and}
  \bibinfo{author}{\bibfnamefont{A.}~\bibnamefont{Kitaev}},
  \bibinfo{journal}{Phys. Rev. B} \textbf{\bibinfo{volume}{81}},
  \bibinfo{pages}{134509} (\bibinfo{year}{2010}),
  \urlprefix\url{https://link.aps.org/doi/10.1103/PhysRevB.81.134509}.

\bibitem[{\citenamefont{Turner et~al.}(2011)\citenamefont{Turner, Pollmann, and
  Berg}}]{PhysRevB.83.075102}
\bibinfo{author}{\bibfnamefont{A.~M.} \bibnamefont{Turner}},
  \bibinfo{author}{\bibfnamefont{F.}~\bibnamefont{Pollmann}}, \bibnamefont{and}
  \bibinfo{author}{\bibfnamefont{E.}~\bibnamefont{Berg}},
  \bibinfo{journal}{Phys. Rev. B} \textbf{\bibinfo{volume}{83}},
  \bibinfo{pages}{075102} (\bibinfo{year}{2011}),
  \urlprefix\url{https://link.aps.org/doi/10.1103/PhysRevB.83.075102}.

\bibitem[{\citenamefont{Yoshida and Kawakami}(2017)}]{PhysRevB.95.045127}
\bibinfo{author}{\bibfnamefont{T.}~\bibnamefont{Yoshida}} \bibnamefont{and}
  \bibinfo{author}{\bibfnamefont{N.}~\bibnamefont{Kawakami}},
  \bibinfo{journal}{Phys. Rev. B} \textbf{\bibinfo{volume}{95}},
  \bibinfo{pages}{045127} (\bibinfo{year}{2017}),
  \urlprefix\url{https://link.aps.org/doi/10.1103/PhysRevB.95.045127}.

\bibitem[{\citenamefont{Yoshida et~al.}(2017)\citenamefont{Yoshida, Daido,
  Yanase, and Kawakami}}]{PhysRevLett.118.147001}
\bibinfo{author}{\bibfnamefont{T.}~\bibnamefont{Yoshida}},
  \bibinfo{author}{\bibfnamefont{A.}~\bibnamefont{Daido}},
  \bibinfo{author}{\bibfnamefont{Y.}~\bibnamefont{Yanase}}, \bibnamefont{and}
  \bibinfo{author}{\bibfnamefont{N.}~\bibnamefont{Kawakami}},
  \bibinfo{journal}{Phys. Rev. Lett.} \textbf{\bibinfo{volume}{118}},
  \bibinfo{pages}{147001} (\bibinfo{year}{2017}),
  \urlprefix\url{https://link.aps.org/doi/10.1103/PhysRevLett.118.147001}.

\bibitem[{\citenamefont{Dzero et~al.}(2010)\citenamefont{Dzero, Sun, Galitski,
  and Coleman}}]{PhysRevLett.104.106408}
\bibinfo{author}{\bibfnamefont{M.}~\bibnamefont{Dzero}},
  \bibinfo{author}{\bibfnamefont{K.}~\bibnamefont{Sun}},
  \bibinfo{author}{\bibfnamefont{V.}~\bibnamefont{Galitski}}, \bibnamefont{and}
  \bibinfo{author}{\bibfnamefont{P.}~\bibnamefont{Coleman}},
  \bibinfo{journal}{Phys. Rev. Lett.} \textbf{\bibinfo{volume}{104}},
  \bibinfo{pages}{106408} (\bibinfo{year}{2010}),
  \urlprefix\url{http://link.aps.org/doi/10.1103/PhysRevLett.104.106408}.

\bibitem[{\citenamefont{Dzero et~al.}(2012)\citenamefont{Dzero, Sun, Coleman,
  and Galitski}}]{PhysRevB.85.045130}
\bibinfo{author}{\bibfnamefont{M.}~\bibnamefont{Dzero}},
  \bibinfo{author}{\bibfnamefont{K.}~\bibnamefont{Sun}},
  \bibinfo{author}{\bibfnamefont{P.}~\bibnamefont{Coleman}}, \bibnamefont{and}
  \bibinfo{author}{\bibfnamefont{V.}~\bibnamefont{Galitski}},
  \bibinfo{journal}{Phys. Rev. B} \textbf{\bibinfo{volume}{85}},
  \bibinfo{pages}{045130} (\bibinfo{year}{2012}),
  \urlprefix\url{http://link.aps.org/doi/10.1103/PhysRevB.85.045130}.

\bibitem[{\citenamefont{Dzero et~al.}(2016)\citenamefont{Dzero, Xia, Galitski,
  and Coleman}}]{annurev-conmatphys-031214-014749}
\bibinfo{author}{\bibfnamefont{M.}~\bibnamefont{Dzero}},
  \bibinfo{author}{\bibfnamefont{J.}~\bibnamefont{Xia}},
  \bibinfo{author}{\bibfnamefont{V.}~\bibnamefont{Galitski}}, \bibnamefont{and}
  \bibinfo{author}{\bibfnamefont{P.}~\bibnamefont{Coleman}},
  \bibinfo{journal}{Annual Review of Condensed Matter Physics}
  \textbf{\bibinfo{volume}{7}}, \bibinfo{pages}{249} (\bibinfo{year}{2016}),
  \eprint{https://doi.org/10.1146/annurev-conmatphys-031214-014749},
  \urlprefix\url{https://doi.org/10.1146/annurev-conmatphys-031214-014749}.

\bibitem[{\citenamefont{Takimoto}(2011)}]{Takimoto2011}
\bibinfo{author}{\bibfnamefont{T.}~\bibnamefont{Takimoto}},
  \bibinfo{journal}{Journal of the Physical Society of Japan}
  \textbf{\bibinfo{volume}{80}}, \bibinfo{pages}{123710}
  (\bibinfo{year}{2011}), \eprint{http://dx.doi.org/10.1143/JPSJ.80.123710},
  \urlprefix\url{http://dx.doi.org/10.1143/JPSJ.80.123710}.

\bibitem[{\citenamefont{Lu et~al.}(2013)\citenamefont{Lu, Zhao, Weng, Fang, and
  Dai}}]{PhysRevLett.110.096401}
\bibinfo{author}{\bibfnamefont{F.}~\bibnamefont{Lu}},
  \bibinfo{author}{\bibfnamefont{J.}~\bibnamefont{Zhao}},
  \bibinfo{author}{\bibfnamefont{H.}~\bibnamefont{Weng}},
  \bibinfo{author}{\bibfnamefont{Z.}~\bibnamefont{Fang}}, \bibnamefont{and}
  \bibinfo{author}{\bibfnamefont{X.}~\bibnamefont{Dai}},
  \bibinfo{journal}{Phys. Rev. Lett.} \textbf{\bibinfo{volume}{110}},
  \bibinfo{pages}{096401} (\bibinfo{year}{2013}),
  \urlprefix\url{http://link.aps.org/doi/10.1103/PhysRevLett.110.096401}.

\bibitem[{\citenamefont{Legner et~al.}(2015)\citenamefont{Legner, R\"uegg, and
  Sigrist}}]{PhysRevLett.115.156405}
\bibinfo{author}{\bibfnamefont{M.}~\bibnamefont{Legner}},
  \bibinfo{author}{\bibfnamefont{A.}~\bibnamefont{R\"uegg}}, \bibnamefont{and}
  \bibinfo{author}{\bibfnamefont{M.}~\bibnamefont{Sigrist}},
  \bibinfo{journal}{Phys. Rev. Lett.} \textbf{\bibinfo{volume}{115}},
  \bibinfo{pages}{156405} (\bibinfo{year}{2015}),
  \urlprefix\url{https://link.aps.org/doi/10.1103/PhysRevLett.115.156405}.

\bibitem[{\citenamefont{Jiang et~al.}(2013)\citenamefont{Jiang, Li, Zhang, Sun,
  Chen, Ye, Xu, Ge, Tan, Niu et~al.}}]{Jiang2013}
\bibinfo{author}{\bibfnamefont{J.}~\bibnamefont{Jiang}},
  \bibinfo{author}{\bibfnamefont{S.}~\bibnamefont{Li}},
  \bibinfo{author}{\bibfnamefont{T.}~\bibnamefont{Zhang}},
  \bibinfo{author}{\bibfnamefont{Z.}~\bibnamefont{Sun}},
  \bibinfo{author}{\bibfnamefont{F.}~\bibnamefont{Chen}},
  \bibinfo{author}{\bibfnamefont{Z.~R.} \bibnamefont{Ye}},
  \bibinfo{author}{\bibfnamefont{M.}~\bibnamefont{Xu}},
  \bibinfo{author}{\bibfnamefont{Q.~Q.} \bibnamefont{Ge}},
  \bibinfo{author}{\bibfnamefont{S.~Y.} \bibnamefont{Tan}},
  \bibinfo{author}{\bibfnamefont{X.~H.} \bibnamefont{Niu}},
  \bibnamefont{et~al.}, \bibinfo{journal}{Nat Commun}
  \textbf{\bibinfo{volume}{4}} (\bibinfo{year}{2013}), \bibinfo{note}{article},
  \urlprefix\url{http://dx.doi.org/10.1038/ncomms4010}.

\bibitem[{\citenamefont{Neupane et~al.}(2013)\citenamefont{Neupane, Alidoust,
  Xu, Kondo, Ishida, Kim, Liu, Belopolski, Jo, Chang et~al.}}]{Neupane2013}
\bibinfo{author}{\bibfnamefont{M.}~\bibnamefont{Neupane}},
  \bibinfo{author}{\bibfnamefont{N.}~\bibnamefont{Alidoust}},
  \bibinfo{author}{\bibfnamefont{S.-Y.} \bibnamefont{Xu}},
  \bibinfo{author}{\bibfnamefont{T.}~\bibnamefont{Kondo}},
  \bibinfo{author}{\bibfnamefont{Y.}~\bibnamefont{Ishida}},
  \bibinfo{author}{\bibfnamefont{D.~J.} \bibnamefont{Kim}},
  \bibinfo{author}{\bibfnamefont{C.}~\bibnamefont{Liu}},
  \bibinfo{author}{\bibfnamefont{I.}~\bibnamefont{Belopolski}},
  \bibinfo{author}{\bibfnamefont{Y.~J.} \bibnamefont{Jo}},
  \bibinfo{author}{\bibfnamefont{T.-R.} \bibnamefont{Chang}},
  \bibnamefont{et~al.}, \bibinfo{journal}{Nat Commun}
  \textbf{\bibinfo{volume}{4}} (\bibinfo{year}{2013}),
  \urlprefix\url{http://dx.doi.org/10.1038/ncomms3991}.

\bibitem[{\citenamefont{Xu et~al.}(2013)\citenamefont{Xu, Shi, Biswas, Matt,
  Dhaka, Huang, Plumb, Radovi\ifmmode~\acute{c}\else \'{c}\fi{}, Dil,
  Pomjakushina et~al.}}]{PhysRevB.88.121102}
\bibinfo{author}{\bibfnamefont{N.}~\bibnamefont{Xu}},
  \bibinfo{author}{\bibfnamefont{X.}~\bibnamefont{Shi}},
  \bibinfo{author}{\bibfnamefont{P.~K.} \bibnamefont{Biswas}},
  \bibinfo{author}{\bibfnamefont{C.~E.} \bibnamefont{Matt}},
  \bibinfo{author}{\bibfnamefont{R.~S.} \bibnamefont{Dhaka}},
  \bibinfo{author}{\bibfnamefont{Y.}~\bibnamefont{Huang}},
  \bibinfo{author}{\bibfnamefont{N.~C.} \bibnamefont{Plumb}},
  \bibinfo{author}{\bibfnamefont{M.}~\bibnamefont{Radovi\ifmmode~\acute{c}\else
  \'{c}\fi{}}}, \bibinfo{author}{\bibfnamefont{J.~H.} \bibnamefont{Dil}},
  \bibinfo{author}{\bibfnamefont{E.}~\bibnamefont{Pomjakushina}},
  \bibnamefont{et~al.}, \bibinfo{journal}{Phys. Rev. B}
  \textbf{\bibinfo{volume}{88}}, \bibinfo{pages}{121102}
  (\bibinfo{year}{2013}),
  \urlprefix\url{http://link.aps.org/doi/10.1103/PhysRevB.88.121102}.

\bibitem[{\citenamefont{Zhu et~al.}(2013)\citenamefont{Zhu, Nicolaou, Levy,
  Butch, Syers, Wang, Paglione, Sawatzky, Elfimov, and
  Damascelli}}]{PhysRevLett.111.216402}
\bibinfo{author}{\bibfnamefont{Z.-H.} \bibnamefont{Zhu}},
  \bibinfo{author}{\bibfnamefont{A.}~\bibnamefont{Nicolaou}},
  \bibinfo{author}{\bibfnamefont{G.}~\bibnamefont{Levy}},
  \bibinfo{author}{\bibfnamefont{N.~P.} \bibnamefont{Butch}},
  \bibinfo{author}{\bibfnamefont{P.}~\bibnamefont{Syers}},
  \bibinfo{author}{\bibfnamefont{X.~F.} \bibnamefont{Wang}},
  \bibinfo{author}{\bibfnamefont{J.}~\bibnamefont{Paglione}},
  \bibinfo{author}{\bibfnamefont{G.~A.} \bibnamefont{Sawatzky}},
  \bibinfo{author}{\bibfnamefont{I.~S.} \bibnamefont{Elfimov}},
  \bibnamefont{and}
  \bibinfo{author}{\bibfnamefont{A.}~\bibnamefont{Damascelli}},
  \bibinfo{journal}{Phys. Rev. Lett.} \textbf{\bibinfo{volume}{111}},
  \bibinfo{pages}{216402} (\bibinfo{year}{2013}),
  \urlprefix\url{http://link.aps.org/doi/10.1103/PhysRevLett.111.216402}.

\bibitem[{\citenamefont{Frantzeskakis et~al.}(2013)\citenamefont{Frantzeskakis,
  de~Jong, Zwartsenberg, Huang, Pan, Zhang, Zhang, Zhang, Bao, Tegus
  et~al.}}]{PhysRevX.3.041024}
\bibinfo{author}{\bibfnamefont{E.}~\bibnamefont{Frantzeskakis}},
  \bibinfo{author}{\bibfnamefont{N.}~\bibnamefont{de~Jong}},
  \bibinfo{author}{\bibfnamefont{B.}~\bibnamefont{Zwartsenberg}},
  \bibinfo{author}{\bibfnamefont{Y.~K.} \bibnamefont{Huang}},
  \bibinfo{author}{\bibfnamefont{Y.}~\bibnamefont{Pan}},
  \bibinfo{author}{\bibfnamefont{X.}~\bibnamefont{Zhang}},
  \bibinfo{author}{\bibfnamefont{J.~X.} \bibnamefont{Zhang}},
  \bibinfo{author}{\bibfnamefont{F.~X.} \bibnamefont{Zhang}},
  \bibinfo{author}{\bibfnamefont{L.~H.} \bibnamefont{Bao}},
  \bibinfo{author}{\bibfnamefont{O.}~\bibnamefont{Tegus}},
  \bibnamefont{et~al.}, \bibinfo{journal}{Phys. Rev. X}
  \textbf{\bibinfo{volume}{3}}, \bibinfo{pages}{041024} (\bibinfo{year}{2013}),
  \urlprefix\url{http://link.aps.org/doi/10.1103/PhysRevX.3.041024}.

\bibitem[{\citenamefont{Xu et~al.}(2014)\citenamefont{Xu, Biswas, Dil, Dhaka,
  Landolt, Muff, Matt, Shi, Plumb, Radovi{\"A}? et~al.}}]{Xu2014}
\bibinfo{author}{\bibfnamefont{N.}~\bibnamefont{Xu}},
  \bibinfo{author}{\bibfnamefont{P.~K.} \bibnamefont{Biswas}},
  \bibinfo{author}{\bibfnamefont{J.~H.} \bibnamefont{Dil}},
  \bibinfo{author}{\bibfnamefont{R.~S.} \bibnamefont{Dhaka}},
  \bibinfo{author}{\bibfnamefont{G.}~\bibnamefont{Landolt}},
  \bibinfo{author}{\bibfnamefont{S.}~\bibnamefont{Muff}},
  \bibinfo{author}{\bibfnamefont{C.~E.} \bibnamefont{Matt}},
  \bibinfo{author}{\bibfnamefont{X.}~\bibnamefont{Shi}},
  \bibinfo{author}{\bibfnamefont{N.~C.} \bibnamefont{Plumb}},
  \bibinfo{author}{\bibfnamefont{M.}~\bibnamefont{Radovi{\"A}?}},
  \bibnamefont{et~al.}, \bibinfo{journal}{Nat Commun}
  \textbf{\bibinfo{volume}{5}} (\bibinfo{year}{2014}), \bibinfo{note}{article},
  \urlprefix\url{http://dx.doi.org/10.1038/ncomms5566}.

\bibitem[{\citenamefont{Hagiwara et~al.}(2016)\citenamefont{Hagiwara, Ohtsubo,
  Matsunami, Ideta, Tanaka, Miyazaki, Rault, F{\`e}vre, Bertran, Taleb-Ibrahimi
  et~al.}}]{Hagiwara2016}
\bibinfo{author}{\bibfnamefont{K.}~\bibnamefont{Hagiwara}},
  \bibinfo{author}{\bibfnamefont{Y.}~\bibnamefont{Ohtsubo}},
  \bibinfo{author}{\bibfnamefont{M.}~\bibnamefont{Matsunami}},
  \bibinfo{author}{\bibfnamefont{S.-i.} \bibnamefont{Ideta}},
  \bibinfo{author}{\bibfnamefont{K.}~\bibnamefont{Tanaka}},
  \bibinfo{author}{\bibfnamefont{H.}~\bibnamefont{Miyazaki}},
  \bibinfo{author}{\bibfnamefont{J.~E.} \bibnamefont{Rault}},
  \bibinfo{author}{\bibfnamefont{P.~L.} \bibnamefont{F{\`e}vre}},
  \bibinfo{author}{\bibfnamefont{F.}~\bibnamefont{Bertran}},
  \bibinfo{author}{\bibfnamefont{A.}~\bibnamefont{Taleb-Ibrahimi}},
  \bibnamefont{et~al.}, \bibinfo{journal}{Nature Communications}
  \textbf{\bibinfo{volume}{7}}, \bibinfo{pages}{12690 EP }
  (\bibinfo{year}{2016}), \bibinfo{note}{article},
  \urlprefix\url{http://dx.doi.org/10.1038/ncomms12690}.

\bibitem[{\citenamefont{Weng et~al.}(2014)\citenamefont{Weng, Zhao, Wang, Fang,
  and Dai}}]{PhysRevLett.112.016403}
\bibinfo{author}{\bibfnamefont{H.}~\bibnamefont{Weng}},
  \bibinfo{author}{\bibfnamefont{J.}~\bibnamefont{Zhao}},
  \bibinfo{author}{\bibfnamefont{Z.}~\bibnamefont{Wang}},
  \bibinfo{author}{\bibfnamefont{Z.}~\bibnamefont{Fang}}, \bibnamefont{and}
  \bibinfo{author}{\bibfnamefont{X.}~\bibnamefont{Dai}},
  \bibinfo{journal}{Phys. Rev. Lett.} \textbf{\bibinfo{volume}{112}},
  \bibinfo{pages}{016403} (\bibinfo{year}{2014}),
  \urlprefix\url{https://link.aps.org/doi/10.1103/PhysRevLett.112.016403}.

\bibitem[{\citenamefont{Alexandrov et~al.}(2013)\citenamefont{Alexandrov,
  Dzero, and Coleman}}]{PhysRevLett.111.226403}
\bibinfo{author}{\bibfnamefont{V.}~\bibnamefont{Alexandrov}},
  \bibinfo{author}{\bibfnamefont{M.}~\bibnamefont{Dzero}}, \bibnamefont{and}
  \bibinfo{author}{\bibfnamefont{P.}~\bibnamefont{Coleman}},
  \bibinfo{journal}{Phys. Rev. Lett.} \textbf{\bibinfo{volume}{111}},
  \bibinfo{pages}{226403} (\bibinfo{year}{2013}),
  \urlprefix\url{http://link.aps.org/doi/10.1103/PhysRevLett.111.226403}.

\bibitem[{\citenamefont{Peters et~al.}(2016)\citenamefont{Peters, Yoshida,
  Sakakibara, and Kawakami}}]{PhysRevB.93.235159}
\bibinfo{author}{\bibfnamefont{R.}~\bibnamefont{Peters}},
  \bibinfo{author}{\bibfnamefont{T.}~\bibnamefont{Yoshida}},
  \bibinfo{author}{\bibfnamefont{H.}~\bibnamefont{Sakakibara}},
  \bibnamefont{and} \bibinfo{author}{\bibfnamefont{N.}~\bibnamefont{Kawakami}},
  \bibinfo{journal}{Phys. Rev. B} \textbf{\bibinfo{volume}{93}},
  \bibinfo{pages}{235159} (\bibinfo{year}{2016}),
  \urlprefix\url{https://link.aps.org/doi/10.1103/PhysRevB.93.235159}.

\bibitem[{\citenamefont{Chang et~al.}(2017)\citenamefont{Chang, Erten, and
  Coleman}}]{Chang_Po2017}
\bibinfo{author}{\bibfnamefont{P.-Y.} \bibnamefont{Chang}},
  \bibinfo{author}{\bibfnamefont{O.}~\bibnamefont{Erten}}, \bibnamefont{and}
  \bibinfo{author}{\bibfnamefont{P.}~\bibnamefont{Coleman}},
  \bibinfo{journal}{Nature Physics} \textbf{\bibinfo{volume}{13}},
  \bibinfo{pages}{794 EP } (\bibinfo{year}{2017}), \bibinfo{note}{article},
  \urlprefix\url{http://dx.doi.org/10.1038/nphys4092}.

\bibitem[{\citenamefont{Tan et~al.}(2015)\citenamefont{Tan, Hsu, Zeng, Hatnean,
  Harrison, Zhu, Hartstein, Kiourlappou, Srivastava, Johannes et~al.}}]{Tan287}
\bibinfo{author}{\bibfnamefont{B.~S.} \bibnamefont{Tan}},
  \bibinfo{author}{\bibfnamefont{Y.-T.} \bibnamefont{Hsu}},
  \bibinfo{author}{\bibfnamefont{B.}~\bibnamefont{Zeng}},
  \bibinfo{author}{\bibfnamefont{M.~C.} \bibnamefont{Hatnean}},
  \bibinfo{author}{\bibfnamefont{N.}~\bibnamefont{Harrison}},
  \bibinfo{author}{\bibfnamefont{Z.}~\bibnamefont{Zhu}},
  \bibinfo{author}{\bibfnamefont{M.}~\bibnamefont{Hartstein}},
  \bibinfo{author}{\bibfnamefont{M.}~\bibnamefont{Kiourlappou}},
  \bibinfo{author}{\bibfnamefont{A.}~\bibnamefont{Srivastava}},
  \bibinfo{author}{\bibfnamefont{M.~D.} \bibnamefont{Johannes}},
  \bibnamefont{et~al.}, \bibinfo{journal}{Science}
  \textbf{\bibinfo{volume}{349}}, \bibinfo{pages}{287} (\bibinfo{year}{2015}),
  ISSN \bibinfo{issn}{0036-8075},
  \eprint{http://science.sciencemag.org/content/349/6245/287.full.pdf},
  \urlprefix\url{http://science.sciencemag.org/content/349/6245/287}.

\bibitem[{\citenamefont{Belopolski et~al.}(2017)\citenamefont{Belopolski,
  Sanchez, Chang, Manna, Ernst, Xu, Zhang, Zheng, Yin, Singh
  et~al.}}]{Belopolski2015}
\bibinfo{author}{\bibfnamefont{I.}~\bibnamefont{Belopolski}},
  \bibinfo{author}{\bibfnamefont{D.~S.} \bibnamefont{Sanchez}},
  \bibinfo{author}{\bibfnamefont{G.}~\bibnamefont{Chang}},
  \bibinfo{author}{\bibfnamefont{K.}~\bibnamefont{Manna}},
  \bibinfo{author}{\bibfnamefont{B.}~\bibnamefont{Ernst}},
  \bibinfo{author}{\bibfnamefont{S.-Y.} \bibnamefont{Xu}},
  \bibinfo{author}{\bibfnamefont{S.~S.} \bibnamefont{Zhang}},
  \bibinfo{author}{\bibfnamefont{H.}~\bibnamefont{Zheng}},
  \bibinfo{author}{\bibfnamefont{J.}~\bibnamefont{Yin}},
  \bibinfo{author}{\bibfnamefont{B.}~\bibnamefont{Singh}}, \bibnamefont{et~al.}
  (\bibinfo{year}{2017}), \eprint{arXiv:1712.09992}.

\bibitem[{\citenamefont{Barla et~al.}(2005)\citenamefont{Barla, Derr, Sanchez,
  Salce, Lapertot, Doyle, R\"uffer, Lengsdorf, Abd-Elmeguid, and
  Flouquet}}]{PhysRevLett.94.166401}
\bibinfo{author}{\bibfnamefont{A.}~\bibnamefont{Barla}},
  \bibinfo{author}{\bibfnamefont{J.}~\bibnamefont{Derr}},
  \bibinfo{author}{\bibfnamefont{J.~P.} \bibnamefont{Sanchez}},
  \bibinfo{author}{\bibfnamefont{B.}~\bibnamefont{Salce}},
  \bibinfo{author}{\bibfnamefont{G.}~\bibnamefont{Lapertot}},
  \bibinfo{author}{\bibfnamefont{B.~P.} \bibnamefont{Doyle}},
  \bibinfo{author}{\bibfnamefont{R.}~\bibnamefont{R\"uffer}},
  \bibinfo{author}{\bibfnamefont{R.}~\bibnamefont{Lengsdorf}},
  \bibinfo{author}{\bibfnamefont{M.~M.} \bibnamefont{Abd-Elmeguid}},
  \bibnamefont{and} \bibinfo{author}{\bibfnamefont{J.}~\bibnamefont{Flouquet}},
  \bibinfo{journal}{Phys. Rev. Lett.} \textbf{\bibinfo{volume}{94}},
  \bibinfo{pages}{166401} (\bibinfo{year}{2005}),
  \urlprefix\url{https://link.aps.org/doi/10.1103/PhysRevLett.94.166401}.

\bibitem[{\citenamefont{Derr et~al.}(2008)\citenamefont{Derr, Knebel,
  Braithwaite, Salce, Flouquet, Flachbart, Gab\'ani, and
  Shitsevalova}}]{PhysRevB.77.193107}
\bibinfo{author}{\bibfnamefont{J.}~\bibnamefont{Derr}},
  \bibinfo{author}{\bibfnamefont{G.}~\bibnamefont{Knebel}},
  \bibinfo{author}{\bibfnamefont{D.}~\bibnamefont{Braithwaite}},
  \bibinfo{author}{\bibfnamefont{B.}~\bibnamefont{Salce}},
  \bibinfo{author}{\bibfnamefont{J.}~\bibnamefont{Flouquet}},
  \bibinfo{author}{\bibfnamefont{K.}~\bibnamefont{Flachbart}},
  \bibinfo{author}{\bibfnamefont{S.}~\bibnamefont{Gab\'ani}}, \bibnamefont{and}
  \bibinfo{author}{\bibfnamefont{N.}~\bibnamefont{Shitsevalova}},
  \bibinfo{journal}{Phys. Rev. B} \textbf{\bibinfo{volume}{77}},
  \bibinfo{pages}{193107} (\bibinfo{year}{2008}),
  \urlprefix\url{https://link.aps.org/doi/10.1103/PhysRevB.77.193107}.

\bibitem[{\citenamefont{Nishiyama et~al.}(2013)\citenamefont{Nishiyama, Mito,
  Pristáš, Hara, Koyama, Ueda, Kohara, Akahama, Gabáni, Reiffers
  et~al.}}]{JPSJ.82.123707}
\bibinfo{author}{\bibfnamefont{K.}~\bibnamefont{Nishiyama}},
  \bibinfo{author}{\bibfnamefont{T.}~\bibnamefont{Mito}},
  \bibinfo{author}{\bibfnamefont{G.}~\bibnamefont{Pristáš}},
  \bibinfo{author}{\bibfnamefont{Y.}~\bibnamefont{Hara}},
  \bibinfo{author}{\bibfnamefont{T.}~\bibnamefont{Koyama}},
  \bibinfo{author}{\bibfnamefont{K.}~\bibnamefont{Ueda}},
  \bibinfo{author}{\bibfnamefont{T.}~\bibnamefont{Kohara}},
  \bibinfo{author}{\bibfnamefont{Y.}~\bibnamefont{Akahama}},
  \bibinfo{author}{\bibfnamefont{S.}~\bibnamefont{Gabáni}},
  \bibinfo{author}{\bibfnamefont{M.}~\bibnamefont{Reiffers}},
  \bibnamefont{et~al.}, \bibinfo{journal}{Journal of the Physical Society of
  Japan} \textbf{\bibinfo{volume}{82}}, \bibinfo{pages}{123707}
  (\bibinfo{year}{2013}), \eprint{https://doi.org/10.7566/JPSJ.82.123707},
  \urlprefix\url{https://doi.org/10.7566/JPSJ.82.123707}.

\bibitem[{\citenamefont{Butch et~al.}(2016)\citenamefont{Butch, Paglione, Chow,
  Xiao, Marianetti, Booth, and Jeffries}}]{PhysRevLett.116.156401}
\bibinfo{author}{\bibfnamefont{N.~P.} \bibnamefont{Butch}},
  \bibinfo{author}{\bibfnamefont{J.}~\bibnamefont{Paglione}},
  \bibinfo{author}{\bibfnamefont{P.}~\bibnamefont{Chow}},
  \bibinfo{author}{\bibfnamefont{Y.}~\bibnamefont{Xiao}},
  \bibinfo{author}{\bibfnamefont{C.~A.} \bibnamefont{Marianetti}},
  \bibinfo{author}{\bibfnamefont{C.~H.} \bibnamefont{Booth}}, \bibnamefont{and}
  \bibinfo{author}{\bibfnamefont{J.~R.} \bibnamefont{Jeffries}},
  \bibinfo{journal}{Phys. Rev. Lett.} \textbf{\bibinfo{volume}{116}},
  \bibinfo{pages}{156401} (\bibinfo{year}{2016}),
  \urlprefix\url{https://link.aps.org/doi/10.1103/PhysRevLett.116.156401}.

\bibitem[{\citenamefont{Chang and Chen}(2017)}]{Chang2017}
\bibinfo{author}{\bibfnamefont{K.-W.} \bibnamefont{Chang}} \bibnamefont{and}
  \bibinfo{author}{\bibfnamefont{P.-J.} \bibnamefont{Chen}}
  (\bibinfo{year}{2017}), \eprint{arXiv:1710.10423}.

\bibitem[{\citenamefont{Fu and Kane}(2007)}]{PhysRevB.76.045302}
\bibinfo{author}{\bibfnamefont{L.}~\bibnamefont{Fu}} \bibnamefont{and}
  \bibinfo{author}{\bibfnamefont{C.~L.} \bibnamefont{Kane}},
  \bibinfo{journal}{Phys. Rev. B} \textbf{\bibinfo{volume}{76}},
  \bibinfo{pages}{045302} (\bibinfo{year}{2007}),
  \urlprefix\url{http://link.aps.org/doi/10.1103/PhysRevB.76.045302}.

\bibitem[{\citenamefont{Fu et~al.}(2007)\citenamefont{Fu, Kane, and
  Mele}}]{PhysRevLett.98.106803}
\bibinfo{author}{\bibfnamefont{L.}~\bibnamefont{Fu}},
  \bibinfo{author}{\bibfnamefont{C.~L.} \bibnamefont{Kane}}, \bibnamefont{and}
  \bibinfo{author}{\bibfnamefont{E.~J.} \bibnamefont{Mele}},
  \bibinfo{journal}{Phys. Rev. Lett.} \textbf{\bibinfo{volume}{98}},
  \bibinfo{pages}{106803} (\bibinfo{year}{2007}),
  \urlprefix\url{http://link.aps.org/doi/10.1103/PhysRevLett.98.106803}.

\bibitem[{\citenamefont{Wang and Zhang}(2012)}]{PhysRevX.2.031008}
\bibinfo{author}{\bibfnamefont{Z.}~\bibnamefont{Wang}} \bibnamefont{and}
  \bibinfo{author}{\bibfnamefont{S.-C.} \bibnamefont{Zhang}},
  \bibinfo{journal}{Phys. Rev. X} \textbf{\bibinfo{volume}{2}},
  \bibinfo{pages}{031008} (\bibinfo{year}{2012}),
  \urlprefix\url{http://link.aps.org/doi/10.1103/PhysRevX.2.031008}.

\bibitem[{\citenamefont{Georges et~al.}(1996)\citenamefont{Georges, Kotliar,
  Krauth, and Rozenberg}}]{Georges1996}
\bibinfo{author}{\bibfnamefont{A.}~\bibnamefont{Georges}},
  \bibinfo{author}{\bibfnamefont{G.}~\bibnamefont{Kotliar}},
  \bibinfo{author}{\bibfnamefont{W.}~\bibnamefont{Krauth}}, \bibnamefont{and}
  \bibinfo{author}{\bibfnamefont{M.~J.} \bibnamefont{Rozenberg}},
  \bibinfo{journal}{Rev. Mod. Phys.} \textbf{\bibinfo{volume}{68}},
  \bibinfo{pages}{13} (\bibinfo{year}{1996}),
  \urlprefix\url{http://link.aps.org/doi/10.1103/RevModPhys.68.13}.

\bibitem[{\citenamefont{Bulla et~al.}(2008)\citenamefont{Bulla, Costi, and
  Pruschke}}]{Bulla2008}
\bibinfo{author}{\bibfnamefont{R.}~\bibnamefont{Bulla}},
  \bibinfo{author}{\bibfnamefont{T.~A.} \bibnamefont{Costi}}, \bibnamefont{and}
  \bibinfo{author}{\bibfnamefont{T.}~\bibnamefont{Pruschke}},
  \bibinfo{journal}{Rev. Mod. Phys.} \textbf{\bibinfo{volume}{80}},
  \bibinfo{pages}{395} (\bibinfo{year}{2008}),
  \urlprefix\url{http://link.aps.org/doi/10.1103/RevModPhys.80.395}.

\bibitem[{\citenamefont{Peters et~al.}(2006)\citenamefont{Peters, Pruschke, and
  Anders}}]{Peters2006}
\bibinfo{author}{\bibfnamefont{R.}~\bibnamefont{Peters}},
  \bibinfo{author}{\bibfnamefont{T.}~\bibnamefont{Pruschke}}, \bibnamefont{and}
  \bibinfo{author}{\bibfnamefont{F.~B.} \bibnamefont{Anders}},
  \bibinfo{journal}{Phys. Rev. B} \textbf{\bibinfo{volume}{74}},
  \bibinfo{pages}{245114} (\bibinfo{year}{2006}),
  \urlprefix\url{http://link.aps.org/doi/10.1103/PhysRevB.74.245114}.

\bibitem[{\citenamefont{Weichselbaum and von Delft}(2007)}]{Weichselbaum2007}
\bibinfo{author}{\bibfnamefont{A.}~\bibnamefont{Weichselbaum}}
  \bibnamefont{and} \bibinfo{author}{\bibfnamefont{J.}~\bibnamefont{von
  Delft}}, \bibinfo{journal}{Phys. Rev. Lett.} \textbf{\bibinfo{volume}{99}},
  \bibinfo{pages}{076402} (\bibinfo{year}{2007}),
  \urlprefix\url{http://link.aps.org/doi/10.1103/PhysRevLett.99.076402}.

\bibitem[{\citenamefont{Doniach}(1977)}]{doniach77}
\bibinfo{author}{\bibfnamefont{S.}~\bibnamefont{Doniach}},
  \bibinfo{journal}{Physica B+C} \textbf{\bibinfo{volume}{91}},
  \bibinfo{pages}{231 } (\bibinfo{year}{1977}), ISSN \bibinfo{issn}{0378-4363},
  \urlprefix\url{http://www.sciencedirect.com/science/article/pii/0378436377901905}.

\bibitem[{\citenamefont{Peters and Kawakami}(2015)}]{PhysRevB.92.075103}
\bibinfo{author}{\bibfnamefont{R.}~\bibnamefont{Peters}} \bibnamefont{and}
  \bibinfo{author}{\bibfnamefont{N.}~\bibnamefont{Kawakami}},
  \bibinfo{journal}{Phys. Rev. B} \textbf{\bibinfo{volume}{92}},
  \bibinfo{pages}{075103} (\bibinfo{year}{2015}),
  \urlprefix\url{https://link.aps.org/doi/10.1103/PhysRevB.92.075103}.

\bibitem[{\citenamefont{Peters and Kawakami}(2017)}]{PhysRevB.96.115158}
\bibinfo{author}{\bibfnamefont{R.}~\bibnamefont{Peters}} \bibnamefont{and}
  \bibinfo{author}{\bibfnamefont{N.}~\bibnamefont{Kawakami}},
  \bibinfo{journal}{Phys. Rev. B} \textbf{\bibinfo{volume}{96}},
  \bibinfo{pages}{115158} (\bibinfo{year}{2017}),
  \urlprefix\url{https://link.aps.org/doi/10.1103/PhysRevB.96.115158}.

\bibitem[{\citenamefont{Beach and Assaad}(2008)}]{PhysRevB.77.205123}
\bibinfo{author}{\bibfnamefont{K.~S.~D.} \bibnamefont{Beach}} \bibnamefont{and}
  \bibinfo{author}{\bibfnamefont{F.~F.} \bibnamefont{Assaad}},
  \bibinfo{journal}{Phys. Rev. B} \textbf{\bibinfo{volume}{77}},
  \bibinfo{pages}{205123} (\bibinfo{year}{2008}),
  \urlprefix\url{https://link.aps.org/doi/10.1103/PhysRevB.77.205123}.

\bibitem[{\citenamefont{Viola~Kusminskiy
  et~al.}(2008)\citenamefont{Viola~Kusminskiy, Beach, Castro~Neto, and
  Campbell}}]{PhysRevB.77.094419}
\bibinfo{author}{\bibfnamefont{S.}~\bibnamefont{Viola~Kusminskiy}},
  \bibinfo{author}{\bibfnamefont{K.~S.~D.} \bibnamefont{Beach}},
  \bibinfo{author}{\bibfnamefont{A.~H.} \bibnamefont{Castro~Neto}},
  \bibnamefont{and} \bibinfo{author}{\bibfnamefont{D.~K.}
  \bibnamefont{Campbell}}, \bibinfo{journal}{Phys. Rev. B}
  \textbf{\bibinfo{volume}{77}}, \bibinfo{pages}{094419}
  (\bibinfo{year}{2008}),
  \urlprefix\url{https://link.aps.org/doi/10.1103/PhysRevB.77.094419}.

\bibitem[{\citenamefont{Li et~al.}(2010)\citenamefont{Li, Zhang, and
  Yu}}]{PhysRevB.81.094420}
\bibinfo{author}{\bibfnamefont{G.-B.} \bibnamefont{Li}},
  \bibinfo{author}{\bibfnamefont{G.-M.} \bibnamefont{Zhang}}, \bibnamefont{and}
  \bibinfo{author}{\bibfnamefont{L.}~\bibnamefont{Yu}}, \bibinfo{journal}{Phys.
  Rev. B} \textbf{\bibinfo{volume}{81}}, \bibinfo{pages}{094420}
  (\bibinfo{year}{2010}),
  \urlprefix\url{https://link.aps.org/doi/10.1103/PhysRevB.81.094420}.

\bibitem[{\citenamefont{Peters et~al.}(2012)\citenamefont{Peters, Kawakami, and
  Pruschke}}]{Peters2012}
\bibinfo{author}{\bibfnamefont{R.}~\bibnamefont{Peters}},
  \bibinfo{author}{\bibfnamefont{N.}~\bibnamefont{Kawakami}}, \bibnamefont{and}
  \bibinfo{author}{\bibfnamefont{T.}~\bibnamefont{Pruschke}},
  \bibinfo{journal}{Phys. Rev. Lett.} \textbf{\bibinfo{volume}{108}},
  \bibinfo{pages}{086402} (\bibinfo{year}{2012}),
  \urlprefix\url{http://link.aps.org/doi/10.1103/PhysRevLett.108.086402}.

\bibitem[{\citenamefont{Irkhin and Katsnelson}(1991)}]{Irkhin1991}
\bibinfo{author}{\bibfnamefont{V.~Y.} \bibnamefont{Irkhin}} \bibnamefont{and}
  \bibinfo{author}{\bibfnamefont{M.~I.} \bibnamefont{Katsnelson}},
  \bibinfo{journal}{Zeitschrift f{\"u}r Physik B Condensed Matter}
  \textbf{\bibinfo{volume}{82}}, \bibinfo{pages}{77} (\bibinfo{year}{1991}),
  ISSN \bibinfo{issn}{1431-584X},
  \urlprefix\url{https://doi.org/10.1007/BF01313989}.

\bibitem[{\citenamefont{Yoshida et~al.}(2013)\citenamefont{Yoshida, Peters,
  Fujimoto, and Kawakami}}]{PhysRevB.87.165109}
\bibinfo{author}{\bibfnamefont{T.}~\bibnamefont{Yoshida}},
  \bibinfo{author}{\bibfnamefont{R.}~\bibnamefont{Peters}},
  \bibinfo{author}{\bibfnamefont{S.}~\bibnamefont{Fujimoto}}, \bibnamefont{and}
  \bibinfo{author}{\bibfnamefont{N.}~\bibnamefont{Kawakami}},
  \bibinfo{journal}{Phys. Rev. B} \textbf{\bibinfo{volume}{87}},
  \bibinfo{pages}{165109} (\bibinfo{year}{2013}),
  \urlprefix\url{http://link.aps.org/doi/10.1103/PhysRevB.87.165109}.

\bibitem[{\citenamefont{Gole\ifmmode~\check{z}\else \v{z}\fi{} and
  \ifmmode~\check{Z}\else \v{Z}\fi{}itko}(2013)}]{PhysRevB.88.054431}
\bibinfo{author}{\bibfnamefont{D.}~\bibnamefont{Gole\ifmmode~\check{z}\else
  \v{z}\fi{}}} \bibnamefont{and}
  \bibinfo{author}{\bibfnamefont{R.}~\bibnamefont{\ifmmode~\check{Z}\else
  \v{Z}\fi{}itko}}, \bibinfo{journal}{Phys. Rev. B}
  \textbf{\bibinfo{volume}{88}}, \bibinfo{pages}{054431}
  (\bibinfo{year}{2013}),
  \urlprefix\url{https://link.aps.org/doi/10.1103/PhysRevB.88.054431}.

\bibitem[{\citenamefont{Fu}(2011)}]{PhysRevLett.106.106802}
\bibinfo{author}{\bibfnamefont{L.}~\bibnamefont{Fu}}, \bibinfo{journal}{Phys.
  Rev. Lett.} \textbf{\bibinfo{volume}{106}}, \bibinfo{pages}{106802}
  (\bibinfo{year}{2011}),
  \urlprefix\url{https://link.aps.org/doi/10.1103/PhysRevLett.106.106802}.

\bibitem[{\citenamefont{Haldane}(2004)}]{PhysRevLett.93.206602}
\bibinfo{author}{\bibfnamefont{F.~D.~M.} \bibnamefont{Haldane}},
  \bibinfo{journal}{Phys. Rev. Lett.} \textbf{\bibinfo{volume}{93}},
  \bibinfo{pages}{206602} (\bibinfo{year}{2004}),
  \urlprefix\url{https://link.aps.org/doi/10.1103/PhysRevLett.93.206602}.

\bibitem[{\citenamefont{Gurarie}(2011)}]{PhysRevB.83.085426}
\bibinfo{author}{\bibfnamefont{V.}~\bibnamefont{Gurarie}},
  \bibinfo{journal}{Phys. Rev. B} \textbf{\bibinfo{volume}{83}},
  \bibinfo{pages}{085426} (\bibinfo{year}{2011}),
  \urlprefix\url{https://link.aps.org/doi/10.1103/PhysRevB.83.085426}.

\bibitem[{\citenamefont{Ishikawa and Matsuyama}(1987)}]{ISHIKAWA1987523}
\bibinfo{author}{\bibfnamefont{K.}~\bibnamefont{Ishikawa}} \bibnamefont{and}
  \bibinfo{author}{\bibfnamefont{T.}~\bibnamefont{Matsuyama}},
  \bibinfo{journal}{Nuclear Physics B} \textbf{\bibinfo{volume}{280}},
  \bibinfo{pages}{523 } (\bibinfo{year}{1987}), ISSN \bibinfo{issn}{0550-3213},
  \urlprefix\url{http://www.sciencedirect.com/science/article/pii/055032138790160X}.

\end{thebibliography}

\end{document}